\documentclass[journal,draftclsnofoot,onecolumn]{IEEEtran}
\usepackage{graphicx}
\usepackage{amssymb}

\usepackage{caption}
\renewcommand{\theequation}{\arabic{section}.\arabic{equation}}
\newtheorem{theorem}{Theorem}

\newtheorem{lemma}{Lemma}
\newtheorem{definition}{Definition}

\newtheorem{remark}{Remark}

\begin{document}

\title{Degraded Broadcast Channel with Side Information, Confidential Messages and with or without Noiseless Feedback}

\author{Bin~Dai,
        A.~J.~Han~Vinck,~\IEEEmembership{Fellow,~IEEE,}
        Zhuojun~Zhuang,
        and~Yuan Luo
\thanks{B. Dai is with the
Computer Science and Engineering Department,
Shanghai Jiao Tong University, and the
Institute for Experimental Mathematics, Duisburg-Essen University,
Ellernstr.29, 45326 Essen, Germany e-mail: daibin007@sjtu.edu.cn.}
\thanks{A. J. Han Vinck is with the
Institute for Experimental Mathematics, Duisburg-Essen University,
Ellernstr.29, 45326 Essen, Germany
e-mail: vinck@iem.uni-due.de.}
\thanks{Z. Zhuang and Y. Luo are with the
Computer Science and Engineering Department,
Shanghai Jiao Tong University,
Shanghai 200240, China e-mail: zhuojunzzj@sjtu.edu.cn, yuanluo@sjtu.edu.cn.}
}

\maketitle

\begin{abstract}
In this paper, first, we investigate the model of degraded broadcast channel with side information and confidential messages.
This work is from Steinberg's work on the degraded broadcast channel with causal and noncausal side information, and
Csisz$\acute{a}$r-K\"{o}rner's work on broadcast channel with confidential messages. Inner and outer bounds on the capacity-equivocation
regions are provided for the noncausal and causal cases. Superposition coding and double-binning technique are used in the
corresponding achievability proofs.

Then, we investigate the degraded broadcast channel with side information, confidential messages and noiseless feedback.
The noiseless feedback is from  the non-degraded receiver to the channel encoder. Inner and outer bounds on the capacity-equivocation region
are provided for the noncausal case, and the capacity-equivocation region is determined for the causal case.
Compared with the model without feedback, we find that the noiseless feedback helps to enlarge the inner bounds for both causal and noncausal cases.
In the achievability proof of the feedback model, the noiseless feedback is used as a secret key shared by the non-degraded receiver and the transmitter, and therefore,
the code construction for the feedback model is a combination of superposition coding, Gel'fand-Pinsker's binning, block Markov coding
and Ahlswede-Cai's secret key on the feedback system.

\end{abstract}


\section{Introduction \label{secI}}
\begin{figure}[htb]
\centering
\includegraphics[scale=0.6]{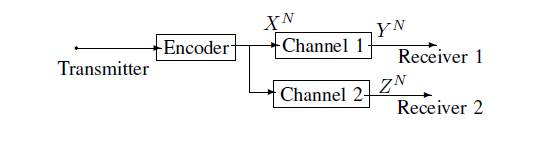}
\caption{Broadcast channels with confidential messages}
\label{f1}
\end{figure}

Equivocation was first introduced into channel coding by Wyner
in his study of wiretap channel \cite{Wy}. It is a kind of degraded broadcast channels.
The object is to transmit messages to the legitimate receiver, while keeping the wiretapper
as ignorant of the messages as possible. After the publication of Wyner's work,
Csisz$\acute{a}$r and K\"{o}rner \cite{CK} investigated a more
general situation: the broadcast channels with confidential
messages, see Figure \ref{f1}. The model of \cite{CK} is to
transmit confidential messages to receiver 1 at rate $R_{1}$
and common messages to both receivers at rate $R_{0}$, while keeping
receiver 2 as ignorant of the confidential messages as possible.
Measuring ignorance by equivocation, a single-letter characterization of
all the achievable triples $(R_{1}, R_{e}, R_{0})$ was provided in \cite{CK},
where $R_{e}$ is the second receiver's equivocation to the confidential messages.
Note that the model of \cite{CK} is also a generalization of \cite{KM},
where no confidentiality condition is imposed. In addition, Merhav \cite{Me} studied a specified wiretap channel, and obtained the capacity region,
where both the legitimate receiver and the wiretapper have access
to some leaked symbols from the source, but the channels for the wiretapper
are more noisy than the legitimate receiver, which shares a secret key with the encoder.

In communication systems there is often a feedback link from the receiver to the transmitter,e.g. the
two-way channels for telephone connections.
It is well known that feedback does not increase the capacity of discrete memoryless channel (DMC).
However, does the feedback increase the capacity region of the wiretap channel?
In order to solve this problem,
Ahlswede and Cai studied the general wiretap channel (the wiretap channel does not need to be degraded)
with noiseless feedback from the legitimate receiver \cite{AC},
and both upper and lower bounds of the secrecy capacity were provided.
Specifically, for the degraded case, they showed that the secrecy capacity is larger than that of Wyner's wiretap channel (without feedback).
In the achievability proof, Ahlswede and Cai \cite{AC} used the noiseless feedback as a
secret key shared by the transmitter and the legitimate receiver,
while the wiretapper had no additional knowledge about the key except his own received symbols.
Besides Ahlswede and Cai's work, the wiretap channel with noisy feedback was studied in \cite{LGP},
and the wiretap channel with secure rate-limited feedback was studied in \cite{AFJK}, and both of them focused on bounds of
the secrecy capacity.

The coding for channels with causal side information at the encoder was first investigated
by Shannon \cite{Sh} in 1958. After that,
in order to solve the problem of coding for a computer memory with
defective cells, Kuznetsov and Tsybakov \cite{KT}
considered a channel in the presence of noncausal side
information at the transmitter. They provided some coding techniques
without determination of the capacity. The capacity was found in
1980 by Gel'fand and Pinsker \cite{GP}. Furthermore, Costa \cite{Co} investigated a power-constrained additive
noise channel, where part of the noise is known at the transmitter
as side information. This channel is also called dirty paper
channel. Based on the dirty paper channel, C. Mitrpant et al. \cite{MVL}
studied  the Gaussian wiretap channel with side information, and provided an inner bound on the capacity-equivocation region. Furthermore,
Y. Chen et al. \cite{Ch} investigated the discrete memoryless wiretap channel with noncausal side information, and also provided an inner
bound on the capacity-equivocation region. Note that the coding scheme of \cite{Ch} is a combination of those in \cite{GP,Wy}.
In order to introduce side information to the broadcast channel,
Steinberg investigated the degraded broadcast channel with side information
\cite{St}, where both causal and noncausal side information were considered
in his paper. Specifically, inner and outer bounds on capacity region were provided for the degraded broadcast
channel with noncausal side information \cite{St},
and meanwhile, the capacity region of the degraded broadcast channel with causal side information was totally determined \cite{St}.

\begin{figure}[htb]
\centering
\includegraphics[scale=0.6]{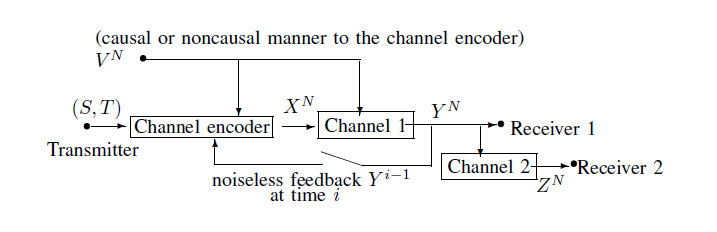}
\caption{Degraded broadcast channel with side information, confidential messages and with or without noiseless feedback}
\label{f2}
\end{figure}

In this paper, we study the model of degraded broadcast channel with side information, confidential messages and with or without
noiseless feedback (see Figure \ref{f2}).
The model of this paper is from Steinberg's work on degraded broadcast channel with side information at the encoder, and
the model of broadcast channel with confidential messages provided by  Csisz$\acute{a}$r and K\"{o}rner.
In Figure \ref{f2}, $S$ is the confidential message sent to receiver 1, and $T$ is the common message sent
to both receiver 1 and receiver 2.
The transition probability of channel 1 depends on
the side information $V^{N}$, and
$V^{N}$ is available to the channel encoder in a causal or noncausal manner. Receiver 2 can get a degraded
version of $Y^{N}$ via channel 2.
In addition, there may exist a noiseless feedback
from the output of channel 1
to the channel encoder.

In this paper, random variab1es, sample values and
alphabets are denoted by capital letters, lower case letters and calligraphic letters, respectively.
A similar convention is applied to the random vectors and their sample values.
\textbf{For example, $U^{N}$ denotes a random $N$-vector $(U_{1},...,U_{N})$,
and $u^{N}=(u_{1},...,u_{N})$ is a specific vector value in $\mathcal{U}^{N}$
that is the $N$th Cartesian power of $\mathcal{U}$.
$U_{i}^{N}$ denotes a random $N-i+1$-vector $(U_{i},...,U_{N})$,
and $u_{i}^{N}=(u_{i},...,u_{N})$ is a specific vector value in $\mathcal{U}_{i}^{N}$.}
Let $p_{V}(v)$ denote the probability mass function $Pr\{V=v\}$.
Throughout the paper, the logarithmic function is to the base 2.

The organization of this paper is as follows.
In Section \ref{secII}, inner and outer bounds on the capacity-equivocation regions
of the model of Figure \ref{f2} without feedback are provided by Theorem \ref{T1} to Theorem \ref{T4}.
Specifically, the inner and outer bounds for the model of Figure \ref{f2} with noncausal side information and without feedback are
provided in Theorem \ref{T1} and Theorem \ref{T2}, respectively.
The inner and outer bounds for the model with causal side information and without feedback are
provided in Theorem \ref{T3} and Theorem \ref{T4}, respectively.
In Section \ref{secIIa}, inner and outer bounds on the capacity-equivocation region of the model of Figure \ref{f2} with noncausal side information
and noiseless feedback are provided in Theorem \ref{T5} and Theorem \ref{T6}, respectively.
The capacity-equivocation region of the model with causal side information and noiseless feedback is given in Theorem \ref{T7}.
Section \ref{secVIII} is for a binary example about the model of Figure \ref{f2} with causal side information and noiseless feedback.
Final conclusions
are provided in Section \ref{secV}.

\section{The Model of Figure \ref{f2} without Feedback\label{secII}}

In this section, the model of Figure \ref{f2} without feedback is considered into two parts.
The model of Figure \ref{f2} with noncausal side information and without feedback
is described in Subsection \ref{sec2.1}, and the model of Figure \ref{f2} with
causal side information and without feedback is described in Subsection \ref{sec2.2}, see the following.

\subsection{The Model of Figure \ref{f2} with Noncausal Side Information and without Feedback\label{sec2.1}}
In this subsection, a description of the  model of Figure \ref{f2} with
noncausal side information and without feedback is given by Definition \ref{def1} to Definition \ref{def5}.
The inner and outer bounds on the capacity-equivocation region $\mathcal{R}^{(n)}$
composed of all achievable $(R_{0}, R_{1}, R_{e})$ triples are
in Theorem \ref{T1} and Theorem \ref{T2}, where the achievable $(R_{0}, R_{1}, R_{e})$ triple is defined in Definition \ref{def6}.

\begin{definition}(\textbf{Channel encoder})\label{def1}
The confidential message $S$ takes values
in $\mathcal{S}$, and the common message $T$ takes values in $\mathcal{T}$. $S$ and $T$ are independent and uniformly
distributed over their ranges.
$V^{N}$ is the side information of channel 1, and it is the output of a discrete memoryless source $p_{V}(v)$.
In addition, $V^{N}$ is available to the channel encoder in a noncausal manner.
Note that $V^{N}$ is independent of $S$ and $T$. At the $i$-th time,
the inputs of the channel encoder are $S$, $T$ and $V^{N}$, while the output is $X_{i}$, i.e.,
the channel encoder is a mapping
\begin{equation}\label{e200}
f_{i}: \mathcal{S}\times \mathcal{T}\times \mathcal{V}^{N}\rightarrow \mathcal{X}_{i},
\end{equation}
where
$f_{i}(s,t,v^{N})=x_{i}\in \mathcal{X}$,
$s\in \mathcal{S}$, $t\in \mathcal{T}$ and $v^{N}\in \mathcal{V}^{N}$.
The transmission rates of the confidential message and the common message are $\frac{\log\parallel \mathcal{S}\parallel}{N}$ and
$\frac{\log\parallel \mathcal{T}\parallel}{N}$, respectively.

\end{definition}

\begin{definition}(\textbf{Channel 1})\label{def2}
Channel 1 is a DMC with finite input alphabet
$\mathcal{X}\times \mathcal{V}$, finite output alphabet $\mathcal{Y}$, and
transition probability $Q_{1}(y|x,v)$, where $x\in \mathcal{X},v\in \mathcal{V},y\in
\mathcal{Y}$. $Q_{1}(y^{N}|x^{N},v^{N})=\prod_{n=1}^{N}Q_{1}(y_{n}|x_{n},v_{n})$.
The inputs of Channel 1 are $X^{N}$ and $V^{N}$, while the output is $Y^{N}$.
\end{definition}

\begin{definition}(\textbf{Channel 2})\label{def3}
Channel 2 is a DMC with finite input alphabet
$\mathcal{Y}$, finite output alphabet $\mathcal{Z}$, and
transition probability $Q_{2}(z|y)$, where $y\in \mathcal{Y},z\in
\mathcal{Z}$. $Q_{2}(z^{N}|y^{N})=\prod_{n=1}^{N}Q_{2}(z_{n}|y_{n})$.
The inputs of Channel 2 is $Y^{N}$, while the output is $Z^{N}$.
Receiver 2's equivocation to the confidential message is defined as
\begin{equation}\label{e201}
\Delta=\frac{1}{N}H(S|Z^{N}).
\end{equation}
The cascade of Channel 1 and Channel 2
is another DMC with transition probability
\begin{equation}\label{e201.1}
Q_{3}(z|x,v)=\sum_{y\in\mathcal{Y}}Q_{W}(z|y)Q_{M}(y|x,v).
\end{equation}

\end{definition}

\begin{definition}(\textbf{Decoder 1})\label{def4}
Decoder 1 is a mapping $f_{D1}: \mathcal{Y}^{N}\rightarrow \mathcal{S}\times \mathcal{T}$,
with input $Y^{N}$ and outputs $\widehat{S}, \widehat{T}$. Let $P_{e1}$ be the error probability of receiver 1
, and it is
defined as $Pr\{(S,T)\neq (\widehat{S},\widehat{T})\}$.
\end{definition}

\begin{definition}(\textbf{Decoder 2})\label{def5}
Decoder 2 is a mapping $f_{D2}: \mathcal{Z}^{N}\rightarrow \mathcal{T}$,
with input $Z^{N}$ and output $\widetilde{T}$. Let $P_{e2}$ be the error probability of receiver 2
, and it is defined as $Pr\{T\neq \widetilde{T}\}$.
\end{definition}

\begin{definition}(\textbf{Achievable $(R_{0}, R_{1}, R_{e})$ triple in the model of Figure \ref{f2} with noncausal side information and without feedback})\label{def6}
A triple $(R_{0}, R_{1}, R_{e})$ (where $R_{0}, R_{1}, R_{e}>0$) is called
achievable if, for any $\epsilon>0$ (where $\epsilon$ is an arbitrary small positive real number
and $\epsilon\rightarrow 0$), there exists a channel
encoder-decoder $(N, \Delta, P_{e1}, P_{e2})$ such that
\begin{equation}\label{e202}
\lim_{N\rightarrow \infty}\frac{\log\parallel \mathcal{T}\parallel}{N}= R_{0} \label{e202a},
\lim_{N\rightarrow \infty}\frac{\log\parallel \mathcal{S}\parallel}{N}= R_{1} \label{e202b}, \lim_{N\rightarrow \infty}\Delta\geq R_{e} \label{e202c},
P_{e1}\leq \epsilon\label{e202d}, P_{e2}\leq \epsilon\label{e202e}.
\end{equation}
\end{definition}

The following Theorem \ref{T1} and Theorem \ref{T2} provide inner and outer bounds on the capacity-equivocation region $\mathcal{R}^{(n)}$
for the model of Figure \ref{f2} with noncausal side information and without feedback,
and they are proved in Appendix \ref{appen1} and Appendix \ref{appen2}.

\begin{theorem}\label{T1}
A single-letter characterization of the region $\mathcal{R}^{(ni)}$ is as follows,
\begin{eqnarray*}
&&\mathcal{R}^{(ni)}=\{(R_{0}, R_{1}, R_{e}): 0\leq R_{e}\leq R_{1},\\
&&R_{0}\leq I(U;Z)-I(U;V), \\
&&R_{1}\leq I(K;Y|U)-I(K;V|U),\\
&&R_{e}\leq I(K;Y|U)-I(K;Z|U)\},
\end{eqnarray*}
where $p_{UKVXYZ}(u,k,v,x,y,z)=p_{Z|Y}(z|y)p_{Y|X,V}(y|x,v)p_{UKXV}(u,k,x,v)$, which implies that
$(U,K)\rightarrow (X,V)\rightarrow Y\rightarrow Z$.

The region $\mathcal{R}^{(ni)}$ satisfies $\mathcal{R}^{(ni)}\subseteq \mathcal{R}^{(n)}$.
\end{theorem}

\begin{remark}\label{R1}
There are some notes on Theorem \ref{T1}, see the following.
\begin{itemize}

\item The region $\mathcal{R}^{(ni)}$ is convex, and the proof is directly obtained by introducing a time sharing random variable into Theorem \ref{T1},
and therefore, we omit the proof here.

\item The ranges of the random variables $U$ and $K$ satisfy
$$\|\mathcal{U}\|\leq \|\mathcal{X}\|\|\mathcal{V}\|+2,$$
$$\|\mathcal{K}\|\leq (\|\mathcal{X}\|\|\mathcal{V}\|+2)^{2}.$$
The proof is in Appendix \ref{appen3}.

\item Without the secrecy parameter $R_{e}$, the region $\mathcal{R}^{(ni)}$ is exactly the same as
the achievable region for the degraded broadcast channel with noncausal side information \cite{St}.

\item The points in $\mathcal{R}^{(n)}$ for which $R_{e}=R_{1}$ are of considerable interest, which imply $H(S)=H(S|Z^{N})$.

\begin{definition}(\textbf{The secrecy capacity $C_{s}^{(n)}$})\label{def7.1}
The secrecy capacity $C_{s}^{(n)}$ of the model of Figure \ref{f2} with noncausal side information and without feedback,
is denoted by
\begin{equation}\label{e203}
C_{s}^{(n)}=\max_{(R_{0}=0,R_{1},R_{e}=R_{1})\in \mathcal{R}^{(n)}}R_{1}.
\end{equation}
\end{definition}

Furthermore, the secrecy capacity $C_{s}^{(n)}$ satisfies
\begin{equation}\label{e204}
C_{s}^{(n)}\geq \max\min\{I(K;Y|U)-I(K;Z|U), I(K;Y|U)-I(K;V|U)\}.
\end{equation}

\begin{IEEEproof}[Proof of (\ref{e204})]
Substituting $R_{e}=R_{1}$ and $R_{0}=0$ into the region $\mathcal{R}^{(ni)}$ in Theorem \ref{T1}, we have
\begin{eqnarray}
R_{1}&\leq& I(K;Y|U)-I(K;V|U),\label{e204.5}\\
R_{1}&\leq& I(K;Y|U)-I(K;Z|U).\label{e206}\\
\end{eqnarray}
Note that the triple $(R_{0}=0, R_{1}=\max\min\{I(K;Y|U)-I(K;Z|U), I(K;Y|U)-I(K;V|U)\}, R_{e}=R_{1})$ is achievable, and therefore,
the secrecy capacity $C_{s}^{(n)}\geq \max\min\{I(K;Y|U)-I(K;Z|U), I(K;Y|U)-I(K;V|U)\}$.
Thus the proof is completed.
\end{IEEEproof}

\end{itemize}
\end{remark}

\begin{theorem}\label{T2}
A single-letter characterization of the region $\mathcal{R}^{(no)}$ is as follows,
\begin{eqnarray*}
&&\mathcal{R}^{(no)}=\{(R_{0}, R_{1}, R_{e}): 0\leq R_{e}\leq R_{1},\\
&&R_{0}\leq I(U;Z)-I(U;V), \\
&&R_{1}\leq I(K;Y|U,A)-I(K;V|U,A),\\
&&R_{0}+R_{1}\leq I(U,A,K;Y)-I(U,A,K;V),\\
&&R_{e}\leq I(K;Y|U,A)-I(K;V|U,A)-I(K;Z|U)+I(K;V|U)\},
\end{eqnarray*}
where $p_{UKAVXYZ}(u,k,a,v,x,y,z)=p_{Z|Y}(z|y)p_{Y|X,V}(y|x,v)p_{UKAXV}(u,k,a,x,v)$, which implies that
$(U,K,A)\rightarrow (X,V)\rightarrow Y\rightarrow Z$.

The region $\mathcal{R}^{(no)}$ satisfies $\mathcal{R}^{(n)}\subseteq \mathcal{R}^{(no)}$.
\end{theorem}

\begin{remark}\label{R2}
There are some notes on Theorem \ref{T2}, see the following.
\begin{itemize}

\item The region $\mathcal{R}^{(no)}$ is convex, and the proof is directly obtained by introducing a time sharing random variable into Theorem \ref{T2},
and therefore, we omit the proof here.

\item The ranges of the random variables $U$ and $K$ satisfy
$$\|\mathcal{U}\|\leq \|\mathcal{X}\|\|\mathcal{V}\|+2,$$
$$\|\mathcal{A}\|\leq (\|\mathcal{X}\|\|\mathcal{V}\|+2)(\|\mathcal{X}\|\|\mathcal{V}\|+1),$$
$$\|\mathcal{K}\|\leq (\|\mathcal{X}\|\|\mathcal{V}\|+2)(\|\mathcal{X}\|\|\mathcal{V}\|+1)(\|\mathcal{X}\|\|\mathcal{V}\|+2)^{2}.$$
The proof is similar to the proof of the second part of Remark \ref{R1}, and it is omitted here.

\item Without the secrecy parameter $R_{e}$, the region $\mathcal{R}^{(no)}$ is exactly the same as
the outer bound for the degraded broadcast channel with noncausal side information \cite{St}.

\end{itemize}
\end{remark}

\subsection{The Model of Figure \ref{f2} with Causal Side Information and without Feedback\label{sec2.2}}
The  model of Figure \ref{f2} with causal side information and without feedback is similar to the model in Subsection \ref{sec2.1},
except that the side information $V^{N}$
in Definition \ref{def1} is known to the channel encoder in a causal manner, i.e.,
at the $i$-th time ($1\leq i \leq N$), the output of the channel encoder $x_{i}=f_{i}(s,t,v^{i})$, where
$v^{i}=(v_{1}, v_{2},..., v_{i})$ and $f_{i}$ is the mapping at time $i$. Note that
$V_{i}$ is independent of $(Y^{i-1},T,V_{i+1}^{N},Z^{i-1})$, where $V_{i+1}^{N}=(V_{i+1}, V_{i+2},..., V_{N})$,
$Y^{i-1}=(Y_{1}, Y_{2},..., Y_{i-1})$ and $Z^{i-1}=(Z_{1}, Z_{2},..., Z_{i-1})$.

The following Theorem \ref{T3} and Theorem \ref{T4} provide inner and outer bounds on the capacity-equivocation region $\mathcal{R}^{(c)}$
for the model of Figure \ref{f2} with causal side information and without feedback,
and they are proved in Appendix \ref{appen4} and Appendix \ref{appen5}.

\begin{theorem}\label{T3}
A single-letter characterization of the region $\mathcal{R}^{(ci)}$ is as follows,
\begin{eqnarray*}
&&\mathcal{R}^{(ci)}=\{(R_{0}, R_{1}, R_{e}): 0\leq R_{e}\leq R_{1},\\
&&R_{0}\leq I(U;Z), \\
&&R_{1}\leq I(K;Y|U),\\
&&R_{e}\leq I(K;Y|U)-I(K;Z|U)\},
\end{eqnarray*}
where $p_{UKVXYZ}(u,k,v,x,y,z)=p_{Z|Y}(z|y)p_{Y|X,V}(y|x,v)p_{X|UKV}(x|u,k,v)p_{UK}(u,k)p_{V}(v)$, which implies that
$(U,K)\rightarrow (X,V)\rightarrow Y\rightarrow Z$ and $V$ is independent of $U$ and $K$.

The region $\mathcal{R}^{(ci)}$ satisfies $\mathcal{R}^{(ci)}\subseteq \mathcal{R}^{(c)}$.
\end{theorem}

\begin{remark}\label{R3}
There are some notes on Theorem \ref{T3}, see the following.

\begin{itemize}
\item The region $\mathcal{R}^{(ci)}$ is convex, and the proof is omitted here.

\item The ranges of the random variables $U$ and $K$ satisfy
$$\|\mathcal{U}\|\leq \|\mathcal{X}\|\|\mathcal{V}\|+1,$$
$$\|\mathcal{K}\|\leq (\|\mathcal{X}\|\|\mathcal{V}\|+1)^{2}.$$
The proof is similar to the proof of the second part of Remark \ref{R1}, and it is omitted here.

\item Without the secrecy parameter $R_{e}$, the region $\mathcal{R}^{(ci)}$ is exactly the same as
the capacity region for the degraded broadcast channel with causal side information \cite{St}.

\item The points in $\mathcal{R}^{(c)}$ for which $R_{e}=R_{1}$ are of considerable interest, which imply $H(S)=H(S|Z^{N})$.

\begin{definition}(\textbf{The secrecy capacity $C_{s}^{(c)}$})\label{def7.1}
The secrecy capacity $C_{s}^{(c)}$ of the model of Figure \ref{f2} with causal side information and without feedback,
is denoted by
\begin{equation}\label{e203x}
C_{s}^{(c)}=\max_{(R_{0}=0,R_{1},R_{e}=R_{1})\in \mathcal{R}^{(c)}}R_{1}.
\end{equation}
\end{definition}

Furthermore, the secrecy capacity $C_{s}^{(c)}$ satisfies
\begin{equation}\label{e204x}
C_{s}^{(c)}\geq \max (I(K;Y|U)-I(K;Z|U)).
\end{equation}

\begin{IEEEproof}[Proof of (\ref{e204x})]
Substituting $R_{e}=R_{1}$ and $R_{0}=0$ into the region $\mathcal{R}^{(ci)}$ in Theorem \ref{T3}, we have
\begin{eqnarray*}
R_{1}&\leq& I(K;Y|U),\\
R_{1}&\leq& I(K;Y|U)-I(K;Z|U).
\end{eqnarray*}
Note that the triple $(R_{0}=0, R_{1}=\max (I(K;Y|U)-I(K;Z|U)), R_{e}=R_{1})$ is achievable, and therefore,
the secrecy capacity $C_{s}^{(c)}\geq \max (I(K;Y|U)-I(K;Z|U))$.
Thus the proof is completed.
\end{IEEEproof}

\end{itemize}
\end{remark}

\begin{theorem}\label{T4}
A single-letter characterization of the region $\mathcal{R}^{(co)}$ is as follows,
\begin{eqnarray*}
&&\mathcal{R}^{(co)}=\{(R_{0}, R_{1}, R_{e}): 0\leq R_{e}\leq R_{1},\\
&&R_{0}\leq I(U;Z), \\
&&R_{1}\leq I(K;Y|U),\\
&&R_{e}\leq I(K;Y|U)-I(K;Z|A)\},
\end{eqnarray*}
where $p_{UKAVXYZ}(u,k,a,v,x,y,z)=p_{Z|Y}(z|y)p_{Y|X,V}(y|x,v)p_{X|UAKV}(x|u,a,k,v)p_{UKA}(u,k,a)p_{V}(v)$, which implies that
$(U,K,A)\rightarrow (X,V)\rightarrow Y\rightarrow Z$ and $V$ is independent of $U$, $K$ and $A$.

The region $\mathcal{R}^{(co)}$ satisfies $\mathcal{R}^{(c)}\subseteq \mathcal{R}^{(co)}$.
\end{theorem}

\begin{remark}\label{R4}
There are some notes on Theorem \ref{T4}, see the following.
\begin{itemize}

\item The region $\mathcal{R}^{(co)}$ is convex, and the proof is omitted here.

\item The ranges of the random variables $U$ and $K$ satisfy
$$\|\mathcal{U}\|\leq \|\mathcal{X}\|\|\mathcal{V}\|+1,$$
$$\|\mathcal{A}\|\leq \|\mathcal{X}\|\|\mathcal{V}\|,$$
$$\|\mathcal{K}\|\leq (\|\mathcal{X}\|\|\mathcal{V}\|+1)^{2}\|\mathcal{X}\|\|\mathcal{V}\|.$$
The proof is similar to the proof of the second part of Remark \ref{R1}, and it is omitted here.

\item Without the secrecy parameter $R_{e}$, the region $\mathcal{R}^{(co)}$ is exactly the same as
the capacity region of the degraded broadcast channel with causal side information \cite{St}.

\end{itemize}
\end{remark}

\section{The Model of Figure \ref{f2} with Feedback\label{secIIa}}

In this section, the model of Figure \ref{f2} with feedback is considered into two parts.
The model of Figure \ref{f2} with noncausal side information and feedback
is described in Subsection \ref{sec2.1a}, and the model of Figure \ref{f2} with
causal side information and feedback is described in Subsection \ref{sec2.2a}, see the following.

\subsection{The Model of Figure \ref{f2} with Noncausal Side Information and Feedback\label{sec2.1a}}

The  model of Figure \ref{f2} with noncausal side information and feedback is similar to the model in Subsection \ref{sec2.1},
except that there is a noiseless feedback from receiver 1 to the channel encoder.
The feedback $Y^{i-1}$ (where $2\leq i\leq N$ and
$Y^{i-1}$ takes values in $\mathcal{Y}^{i-1}$) is the previous $i-1$ time
output of channel 1. At the $i$-th time,
the inputs of the channel encoder are $S$, $T$, $Y^{i-1}$ and $V^{N}$, while the output is $X_{i}$, i.e.,
the channel encoder is a mapping
\begin{equation}\label{e200r}
f_{i}: \mathcal{S}\times \mathcal{T}\times \mathcal{Y}^{i-1}\times \mathcal{V}^{N}\rightarrow \mathcal{X}_{i},
\end{equation}
where
$f_{i}(s,t,y^{i-1},v^{N})=x_{i}\in \mathcal{X}$,
$s\in \mathcal{S}$, $t\in \mathcal{T}$, $y^{i-1}\in \mathcal{Y}^{i-1}$ and $v^{N}\in \mathcal{V}^{N}$.

The following Theorem \ref{T5} and Theorem \ref{T6} provide inner and outer bounds on the capacity-equivocation region $\mathcal{R}^{(nf)}$
for the model of Figure \ref{f2} with noncausal side information and feedback,
and they are proved in Appendix \ref{appen6} and Appendix \ref{appen7}.

\begin{theorem}\label{T5}
A single-letter characterization of the region $\mathcal{R}^{(nfi)}$ is as follows,
\begin{eqnarray*}
&&\mathcal{R}^{(nfi)}=\{(R_{0}, R_{1}, R_{e}): 0\leq R_{e}\leq R_{1},\\
&&R_{0}\leq I(U;Z)-I(U;V), \\
&&R_{1}\leq I(K;Y|U)-I(K;V|U),\\
&&R_{e}\leq H(Y|Z)\},
\end{eqnarray*}
where $p_{UKVXYZ}(u,k,v,x,y,z)=p_{Z|Y}(z|y)p_{Y|X,V}(y|x,v)p_{UKXV}(u,k,x,v)$, which implies that
$(U,K)\rightarrow (X,V)\rightarrow Y\rightarrow Z$.

The region $\mathcal{R}^{(nfi)}$ satisfies $\mathcal{R}^{(nfi)}\subseteq \mathcal{R}^{(nf)}$.
\end{theorem}

\begin{remark}\label{R5}
There are some notes on Theorem \ref{T5}, see the following.
\begin{itemize}

\item The region $\mathcal{R}^{(nfi)}$ is convex, and the proof is omitted here.

\item The ranges of the random variables $U$ and $K$ satisfy
$$\|\mathcal{U}\|\leq \|\mathcal{X}\|\|\mathcal{V}\|+2,$$
$$\|\mathcal{K}\|\leq (\|\mathcal{X}\|\|\mathcal{V}\|+2)(\|\mathcal{X}\|\|\mathcal{V}\|+1).$$
The proof is similar to the proof of the second part of Remark \ref{R1}, and it is omitted here.

\item Without the secrecy parameter $R_{e}$, the region $\mathcal{R}^{(nfi)}$ is exactly the same as
the achievable region for the degraded broadcast channel with noncausal side information \cite{St}.

\item The points in $\mathcal{R}^{(nf)}$ for which $R_{e}=R_{1}$ are of considerable interest, which imply $H(S)=H(S|Z^{N})$.

\begin{definition}(\textbf{The secrecy capacity $C_{s}^{(nf)}$})\label{def7.1}
The secrecy capacity $C_{s}^{(nf)}$ of the model of Figure \ref{f2} with noncausal side information and feedback,
is denoted by
\begin{equation}\label{e203r}
C_{s}^{(nf)}=\max_{(R_{0}=0,R_{1},R_{e}=R_{1})\in \mathcal{R}^{(nf)}}R_{1}.
\end{equation}
\end{definition}

Furthermore, the secrecy capacity $C_{s}^{(nf)}$ satisfies
\begin{equation}\label{e204r}
C_{s}^{(nf)}\geq \max\min\{H(Y|Z), I(K;Y|U)-I(K;V|U)\}.
\end{equation}

\begin{IEEEproof}[Proof of (\ref{e204r})]
Substituting $R_{e}=R_{1}$ and $R_{0}=0$ into the region $\mathcal{R}^{(nfi)}$ in Theorem \ref{T5}, we have
\begin{eqnarray*}
R_{1}&\leq& I(K;Y|U)-I(K;V|U),\label{e204.5r}\\
R_{1}&\leq& H(Y|Z).\label{e206r}\\
\end{eqnarray*}
Note that the triple $(R_{0}=0, R_{1}=\max\min\{H(Y|Z), I(K;Y|U)-I(K;V|U)\}, R_{e}=R_{1})$ is achievable, and therefore,
the secrecy capacity $C_{s}^{(nf)}\geq \max\min\{H(Y|Z), I(K;Y|U)-I(K;V|U)\}$.
Thus the proof is completed.
\end{IEEEproof}

\item Note that the formula $R_{e}\leq I(K;Y|U)-I(K;Z|U)$ of Theorem \ref{T1} can be bounded as follows.
\begin{eqnarray}\label{e204rr}
R_{e}&\leq& I(K;Y|U)-I(K;Z|U)\nonumber\\
&=&H(K|U,Z)-H(K|U,Y)\nonumber\\
&\stackrel{(a)}=&H(K|U,Z)-H(K|U,Y,Z)\nonumber\\
&=&I(K;Y|U,Z)\leq H(Y|U,Z)\leq H(Y|Z),
\end{eqnarray}
where (a) is from $K\rightarrow (U,Y)\rightarrow Z$.

Formula (\ref{e204rr}) implies that the feedback helps to enlarge the region $\mathcal{R}^{(ni)}$ in Theorem \ref{T1}.

\end{itemize}
\end{remark}

\begin{theorem}\label{T6}
A single-letter characterization of the region $\mathcal{R}^{(nfo)}$ is as follows,
\begin{eqnarray*}
&&\mathcal{R}^{(nfo)}=\{(R_{0}, R_{1}, R_{e}): 0\leq R_{e}\leq R_{1},\\
&&R_{0}\leq I(U;Z)-I(U;V), \\
&&R_{1}\leq I(K;Y|U,A)-I(K;V|U,A),\\
&&R_{0}+R_{1}\leq I(U,A,K;Y)-I(U,A,K;V),\\
&&R_{e}\leq H(Y|Z)\},
\end{eqnarray*}
where $p_{UKAVXYZ}(u,k,a,v,x,y,z)=p_{Z|Y}(z|y)p_{Y|X,V}(y|x,v)p_{UKAXV}(u,k,a,x,v)$, which implies that
$(U,K,A)\rightarrow (X,V)\rightarrow Y\rightarrow Z$.

The region $\mathcal{R}^{(nfo)}$ satisfies $\mathcal{R}^{(nf)}\subseteq \mathcal{R}^{(nfo)}$.
\end{theorem}

\begin{remark}\label{R6}
There are some notes on Theorem \ref{T6}, see the following.
\begin{itemize}

\item The region $\mathcal{R}^{(nfo)}$ is convex, and the proof is omitted here.

\item The ranges of the random variables $U$ and $K$ satisfy
$$\|\mathcal{U}\|\leq \|\mathcal{X}\|\|\mathcal{V}\|+2,$$
$$\|\mathcal{A}\|\leq (\|\mathcal{X}\|\|\mathcal{V}\|+2)(\|\mathcal{X}\|\|\mathcal{V}\|+1),$$
$$\|\mathcal{K}\|\leq (\|\mathcal{X}\|\|\mathcal{V}\|+2)(\|\mathcal{X}\|\|\mathcal{V}\|+1)(\|\mathcal{X}\|\|\mathcal{V}\|+2)^{2}.$$
The proof is similar to the proof of the second part of Remark \ref{R1}, and it is omitted here.

\item Without the secrecy parameter $R_{e}$, the region $\mathcal{R}^{(nfo)}$ is exactly the same as
the outer bound for the degraded broadcast channel with noncausal side information \cite{St}.

\end{itemize}
\end{remark}

\subsection{The Model of Figure \ref{f2} with Causal Side Information and Feedback\label{sec2.2a}}

The  model of Figure \ref{f2} with causal side information and feedback is similar to the model in Subsection \ref{sec2.1a},
except that the side information $V^{N}$ is known to the channel encoder in a causal manner, i.e.,
at the $i$-th time ($1\leq i \leq N$), the output of the channel encoder $x_{i}=f_{i}(s,t,y^{i-1},v^{i})$, where
$v^{i}=(v_{1}, v_{2},..., v_{i})$ and $f_{i}$ is the mapping at time $i$.

The following Theorem \ref{T7} provides the capacity-equivocation region $\mathcal{R}^{(cf)}$
for the model of Figure \ref{f2} with causal side information and feedback,
and it is proved in Appendix \ref{appen8}.

\begin{theorem}\label{T7}
A single-letter characterization of the capacity-equivocation region $\mathcal{R}^{(cf)}$ is as follows,
\begin{eqnarray*}
&&\mathcal{R}^{(cf)}=\{(R_{0}, R_{1}, R_{e}): 0\leq R_{e}\leq R_{1},\\
&&R_{0}\leq I(U;Z), \\
&&R_{1}\leq I(K;Y|U),\\
&&R_{e}\leq H(Y|Z)\},
\end{eqnarray*}
where $p_{UKVXYZ}(u,k,v,x,y,z)=p_{Z|Y}(z|y)p_{Y|X,V}(y|x,v)p_{X|UKV}(x|u,k,v)p_{UK}(u,k)p_{V}(v)$, which implies that
$(U,K)\rightarrow (X,V)\rightarrow Y\rightarrow Z$ and $V$ is independent of $U$ and $K$.

\end{theorem}

\begin{remark}\label{R7}

There are some notes on Theorem \ref{T7}, see the following.
\begin{itemize}

\item The region $\mathcal{R}^{(cf)}$ is convex, and the proof is omitted here.

\item The ranges of the random variables $U$ and $K$ satisfy
$$\|\mathcal{U}\|\leq \|\mathcal{X}\|\|\mathcal{V}\|+1,$$
$$\|\mathcal{K}\|\leq (\|\mathcal{X}\|\|\mathcal{V}\|+1)\|\mathcal{X}\|\|\mathcal{V}\|.$$
The proof is similar to the proof of the second part of Remark \ref{R1}, and it is omitted here.

\item Without the secrecy parameter $R_{e}$, the region $\mathcal{R}^{(cf)}$ is exactly the same as
the capacity region for the degraded broadcast channel with causal side information \cite{St}, i.e., \textbf{the noiseless
feedback can not increase the capacity of the degraded broadcast channel with causal side information.}

\item The points in $\mathcal{R}^{(cf)}$ for which $R_{e}=R_{1}$ are of considerable interest, which imply $H(S)=H(S|Z^{N})$.

\begin{definition}(\textbf{The secrecy capacity $C_{s}^{(cf)}$})
The secrecy capacity $C_{s}^{(cf)}$ of the model of Figure \ref{f2} with causal side information and feedback,
is denoted by
\begin{equation}\label{e203xx}
C_{s}^{(cf)}=\max_{(R_{0}=0,R_{1},R_{e}=R_{1})\in \mathcal{R}^{(cf)}}R_{1}.
\end{equation}
\end{definition}

Furthermore, the secrecy capacity $C_{s}^{(cf)}$ satisfies
\begin{equation}\label{e204xx}
C_{s}^{(cf)}=\max\min\{I(K;Y|U), H(Y|Z)\}.
\end{equation}

\begin{IEEEproof}[Proof of (\ref{e204xx})]
Substituting $R_{e}=R_{1}$ and $R_{0}=0$ into the region $\mathcal{R}^{(cf)}$ in Theorem \ref{T7}, we have
\begin{equation}\label{e204.5xx}
R_{1}\leq I(K;Y|U),
\end{equation}
\begin{equation}\label{e206xx}
R_{1}\leq H(Y|Z).
\end{equation}
By using (\ref{e203xx}), (\ref{e204.5xx}) and (\ref{e206xx}),
(\ref{e204xx}) is proved.
\end{IEEEproof}

\item By using the same formula as (\ref{e204rr}), it is easy to see that
the feedback helps to enlarge the region $\mathcal{R}^{(ci)}$ in Theorem \ref{T3}.

\end{itemize}
\end{remark}

\section{A Binary Example for the Model of Figure \ref{f2} with Causal Side Information and Noiseless Feedback\label{secVIII}}
\setcounter{equation}{0}

In this section, we calculate the secrecy capacity of a special case of
the model of Figure \ref{f2} with causal side information and noiseless feedback.

Suppose that the channel state information $V^{N}$ is available at the channel encoder in a casual manner, and the random variable
$V$ is uniformly distributed over $\{0,1\}$, i.e., $p_{V}(0)=p_{V}(1)=\frac{1}{2}$. Meanwhile, the random variables $X$, $Y$ and $Z$
take values in $\{0,1\}$, and channel 2 is a BSC (binary symmetric channel) with crossover probability $q$.
The transition probability of channel 1 is defined as follows:

When $v=0$,
\begin{equation}\label{e801}
p_{Y|X,V}(y|x,v=0)=
\left\{
\begin{array}{ll}
1-p, & \mbox{if}\; y=x,\\
p, & \mbox{otherwise}.
\end{array}
\right.
\end{equation}

When $v=1$,
\begin{equation}\label{e802}
p_{Y|X,V}(y|x,v=1)=
\left\{
\begin{array}{ll}
p, & \mbox{if}\; y=x,\\
1-p, & \mbox{otherwise}.
\end{array}
\right.
\end{equation}

From Remark \ref{R7} we know that the secrecy capacity for the causal case is
\begin{equation}\label{e304.bb1}
C_{s}^{(cf)}=\max\min\{I(K;Y|U), H(Y|Z)\}.
\end{equation}
Since $R_{0}=0$, we have $U=const$, and therefore, $I(K;Y|U)=I(K;Y)$.
The remainder of this example is to calculate the characters $\max I(K;Y)$ and $\max H(Y|Z)$.

Let $K$ take values in $\{0,1\}$. The probability of $K$ is defined as follows.
$p_{K}(0)=\alpha$ and $p_{K}(1)=1-\alpha$.

In addition, define the conditional probability mass function $p_{X|K,V}$ as follows.

$p_{X|K,V}(0|0,0)=\beta_{1}$, $p_{X|K,V}(1|0,0)=1-\beta_{1}$,
$p_{X|K,V}(0|0,1)=\beta_{2}$, $p_{X|K,V}(1|0,1)=1-\beta_{2}$,

$p_{X|K,V}(0|1,0)=\beta_{3}$, $p_{X|K,V}(1|1,0)=1-\beta_{3}$,
$p_{X|K,V}(0|1,1)=\beta_{4}$, $p_{X|K,V}(1|1,1)=1-\beta_{4}$.

The character $I(K;Y)$ depends on the joint probability mass functions $p_{KY}$,
and we have
\begin{eqnarray}\label{e806.abc}
p_{KY}(k,y)&=&\sum_{x,v}p_{KYXV}(k,y,x,v)\nonumber\\
&=&\sum_{x,v}p_{Y|XV}(y|x,v)p_{X|K,V}(x|k,v)p_{K}(k)p_{V}(v).
\end{eqnarray}

Now we calculate $\max I(K;Y)$ and $\max H(Y|Z)$, respectively.

\begin{itemize}

\item (Calculation of $\max I(K;Y)$) Since
\begin{equation}\label{e805}
I(K;Y)=\sum_{k}\sum_{y}p_{KY}(k,y)\log\frac{p_{KY}(k,y)}{p_{K}(k)p_{Y}(y)},
\end{equation}
and
\begin{equation}\label{e805.1}
p_{KY}(0,0)=\frac{\alpha}{2}[1-(\beta_{1}-\beta_{2})(1-2p)],
\end{equation}
\begin{equation}\label{e805.2}
p_{KY}(0,1)=\frac{\alpha}{2}[1+(\beta_{1}-\beta_{2})(1-2p)],
\end{equation}
\begin{equation}\label{e805.3}
p_{KY}(1,0)=\frac{\alpha}{2}[1-(\beta_{3}-\beta_{4})(1-2p)],
\end{equation}
\begin{equation}\label{e805.4}
p_{KY}(1,1)=\frac{\alpha}{2}[1+(\beta_{3}-\beta_{4})(1-2p)].
\end{equation}

Define $\beta_{1}-\beta_{2}=a$ and $\beta_{3}-\beta_{4}=b$, then we have
\begin{equation}\label{e805.5}
I(K;Y)\leq h(1-\alpha-p+2p\alpha)-h(p)\leq 1-h(p),
\end{equation}
where $h(x)=-x\log x-(1-x)\log(1-x)$, $0\leq x\leq 1$, and the ``='' is achieved if $a=1$, $b=-1$ and $\alpha=\frac{1}{2}$.

\item The character $H(Y|Z)$ is calculated as follows.

\begin{eqnarray}\label{e805.11}
H(Y|Z)&=&H(Z|Y)+H(Y)-H(Z)\nonumber\\
&\leq&H(Z|Y)=h(q),
\end{eqnarray}
and therefore, $\max H(Y|Z)=h(q)$. Note that $a=1$, $b=-1$ and $\alpha=\frac{1}{2}$
are also the distributions for $H(Y|Z)=h(q)$.

\end{itemize}

Thus, the secrecy capacity of this special case is
\begin{equation}\label{e805.12}
C_{s}^{(cf)}=\min\{1-h(p), h(q)\}.
\end{equation}

\section{Conclusion\label{secV}}

In this paper, we investigate the model of degraded broadcast channel with side information, confidential messages and with or without noiseless feedback.
This work is from Steinberg's work on the degraded broadcast channel with causal and noncausal side information, and
Csisz$\acute{a}$r and K\"{o}rner's work on broadcast channel with confidential messages.
For the non-feedback model,
inner and outer bounds on the capacity-equivocation regions are provided for both causal and noncausal manners.
Superposition coding, Gel'fand-Pinsker's binning and Wyner's random binning technique are used in the corresponding achievability proofs.

For the feedback model, inner and outer bounds
on the capacity-equivocation region are provided for the noncausal case, and the capacity-equivocation region is determined for the causal case.
In the corresponding achievability proofs, the noiseless feedback is used as a secret key shared by receiver 1 and transmitter, and therefore,
the coding schemes for the achievability proofs are a combination of superposition coding, Gel'fand-Pinsker's binning, block Markov coding
 and Ahlswede-Cai's secret key on the feedback system.

Finally, we give an example on calculating the secrecy capacity of the binary degraded broadcast channel with causal side information,
confidential messages and noiseless feedback.

\section*{Acknowledgement}

The authors would like to thank Professor Yossef Steinberg
for his valuable suggestions to improve this paper.

\renewcommand{\theequation}{\arabic{equation}}
\appendices\section{Proof of Theorem \ref{T1}}\label{appen1}

In this section, we will show that any triple
$(R_{0},R_{1},R_{e})\in \mathcal{R}^{ni}$ is achievable.
Superposition coding, Gel'fand-Pinsker's binning and Wyner's random binning technique are used in the construction of the code-books.

Now the
remainder of this section is organized as follows.
The code construction is
in Subsection \ref{app1.1}. The proof of achievability is given in Subsection \ref{app1.2}.

\subsection{Code Construction\label{app1.1}}

Since $R_{e}\leq I(K;Y|U)-I(K;Z|U)$ and $R_{e}\leq R_{1}\leq I(K;Y|U)-I(K;V|U)$, it is sufficient to show that
the triple $(R_{0},R_{1},R_{e}=I(K;Y|U)-\max(I(K;Z|U), I(K;V|U)))$ is achievable, and note that this implies that
$R_{1}\geq R_{e}=I(K;Y|U)-\max(I(K;Z|U), I(K;V|U))$.

Given a triple $(R_{0},R_{1},R_{e})$, choose a joint probability mass function $p_{U,K,V,X,Y,Z}(u,k,v,x,y,z)$
such that
$$0\leq R_{e}\leq R_{1},$$
$$R_{0}\leq I(U;Z)-I(U;V),$$
$$R_{1}\leq I(K;Y|U)-I(K;V|U),$$
$$R_{e}=I(K;Y|U)-\max(I(K;Z|U), I(K;V|U).$$

The confidential message set $\mathcal{S}$ and the common message set $\mathcal{T}$ satisfy the following conditions:
\begin{equation}\label{eapp1}
\lim_{N\rightarrow \infty}\frac{1}{N}\log\parallel \mathcal{S}\parallel=R_{1}=I(K;Y|U)-I(K;V|U)-\gamma_{1},
\end{equation}
\begin{equation}\label{eapp2}
\lim_{N\rightarrow \infty}\frac{1}{N}\log\parallel \mathcal{T}\parallel= R_{0}=I(U;Z)-I(U;V)-\gamma,
\end{equation}
where $\gamma$ and $\gamma_{1}$ are fixed positive real numbers and
\begin{equation}\label{eapp2.5}
0\leq \gamma_{1}\leq^{(a)} \max(I(K;Z|U), I(K;V|U))-I(K;V|U).
\end{equation}
Note that (a) is from $R_{1}\geq R_{e}=I(K;Y|U)-\max(I(K;Z|U), I(K;V|U))$ and (\ref{eapp1}).
Let $\mathcal{S}=\{1,2,...,2^{NR_{1}}\}$ and $\mathcal{T}=\{1,2,...,2^{NR_{0}}\}$.

Code-book generation:

\begin{itemize}

\item \textbf{(Construction of $U^{N}$)}
Gel'fand-Pinsker's binning technique is used in the construction of $U^{N}$, see Figure \ref{f3}.

Generate $2^{N(I(U;Z)-\epsilon_{1,N})}$ ($\epsilon_{1,N}\rightarrow 0$ as $N\rightarrow \infty$)
i.i.d. sequences $u^{N}$, according to the probability mass function
$p_{U}(u)$. Distribute these sequences at random into $2^{NR_{0}}=2^{N(I(U;Z)-I(U;V)-\gamma)}$ bins such that
each bin contains $2^{N(I(U;V)+\gamma-\epsilon_{1,N})}$ sequences. Index each bin by $i\in \{1,2,...,2^{NR_{0}}\}$.

For a given common message $t$ ($t\in \mathcal{T}$) and side information $v^{N}$, try to find
a sequence in bin $t$ \\ $\{u^{N}(t,1),u^{N}(t,2),...,u^{N}(t,2^{N(I(U;V)+\gamma-\epsilon_{1,N})})\}$
that is jointly typical with $v^{N}$, say $u^{N}(t,i^{*})$, i.e., \\ $(u^{N}(t,i^{*}),v^{N})\in T^{N}_{UV}(\epsilon_{1})$.
If multiple such sequences in bin $t$
exist, choose the one with the smallest $i^{*}$. If no such $i^{*}$ exists, then declare an encoding error.

\begin{figure}[htb]
\centering
\includegraphics[scale=0.6]{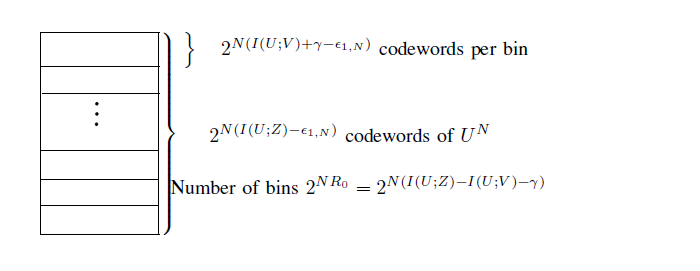}
\caption{Code-book construction for $U^{N}$ in Theorem \ref{T1}}
\label{f3}
\end{figure}

\item \textbf{(Construction of $K^{N}$)}
Classical superposition coding and double binning technique \cite{Ch} are used in the construction of $K^{N}$, see Figure \ref{f4}.

For the transmitted sequence $u^{N}(t,i^{*})$, generate $2^{N(I(K;Y|U)-\epsilon_{2,N})}$
($\epsilon_{2,N}\rightarrow 0$ as $N\rightarrow \infty$) i.i.d. sequences $k^{N}$,
according to the probability mass function $p_{K|U}(k_{i}|u_{i}(t,i^{*}))$.
Distribute these sequences at random into $2^{NR_{1}}=2^{N(I(K;Y|U)-I(K;V|U)-\gamma_{1})}$ bins such that
each bin contains $2^{N(I(K;V|U)+\gamma_{1}-\epsilon_{2,N})}$ sequences. Index each bin by $i\in \{1,2,...,2^{NR_{1}}\}$.
Then place the $2^{N(I(K;V|U)+\gamma_{1}-\epsilon_{2,N})}$ sequences in every bin randomly into
$2^{N(\max(I(K;V|U), I(K;Z|U))-I(K;Z|U)+\epsilon_{3,N})}$ ($\epsilon_{3,N}\rightarrow 0$ as $N\rightarrow \infty$)
subbins such that every subbin contains
$2^{N(I(K;V|U)+\gamma_{1}-\epsilon_{2,N}-\max(I(K;V|U), I(K;Z|U))+I(K;Z|U)-\epsilon_{3,N})}$ sequences.
Let $J$ be the random variable to represent  the index of the subbin. Index each subbin by \\
$j\in \{1,2,...,2^{N(\max(I(K;V|U), I(K;Z|U))-I(K;Z|U)+\epsilon_{3,N})}\}$, i.e.,
\begin{equation}\label{eapp4x}
\log \|\mathcal{J}\|=N(\max(I(K;V|U), I(K;Z|U))-I(K;Z|U)+\epsilon_{3,N}).
\end{equation}

Here note that the number of the sequences in every subbin is upper bounded as follows.
\begin{eqnarray}\label{eapp3}
&&I(K;V|U)+\gamma_{1}-\epsilon_{2,N}-\max(I(K;V|U), I(K;Z|U))+I(K;Z|U)-\epsilon_{3,N}\nonumber\\
&\leq^{(a)}&I(K;Z|U)-\epsilon_{2,N}-\epsilon_{3,N},
\end{eqnarray}
where (a) is from (\ref{eapp2.5}). This implies that
\begin{equation}\label{eapp4}
\lim_{N\rightarrow \infty}H(K^{N}|U^{N},S,J,Z^{N})=0.
\end{equation}

For a given confidential message $s$ ($s\in \mathcal{S}$) and side information $v^{N}$, try to find
a sequence $k^{N}(u^{N}(t,i^{*}))$ in bin $s$ such that  $(k^{N}(u^{N}(t,i^{*})),v^{N})\in T^{N}_{KV|U}(\epsilon_{2})$.
If multiple such sequences in bin $s$
exist, choose the one with the smallest index in the bin. If no such sequence exists, declare an encoding error.

\begin{figure}[htb]
\centering
\includegraphics[scale=0.6]{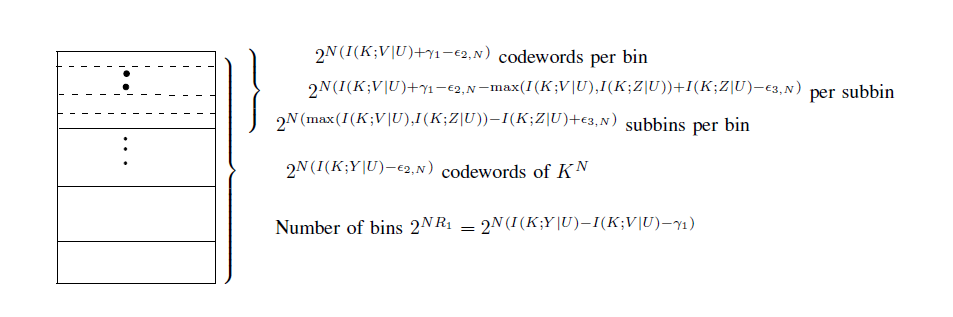}
\caption{Code-book construction for $K^{N}$ in Theorem \ref{T1}}
\label{f4}
\end{figure}

\item \textbf{(Construction of $X^{N}$)}
The $x^{N}$ is generated
according to a new discrete memoryless channel (DMC) with inputs $k^{N}$, $u^{N}$, $v^{N}$,
and output $x^{N}$. The transition probability of this new DMC is $p_{X|U,K,V}(x|u,k,v)$,
which is obtained from the joint probability mass function $p_{U,K,V,X,Y,Z}(u,k,v,x,y,z)$.
The probability \\ $p_{X^{N}|U^{N},K^{N},V^{N}}(x^{N}|u^{N},k^{N},v^{N})$ is calculated as follows.
\begin{equation}\label{eapp4.5}
p_{X^{N}|U^{N},K^{N},V^{N}}(x^{N}|u^{N},k^{N},v^{N})=
\prod_{i=1}^{N}p_{X|U,K,V}(x_{i}|u_{i},k_{i},v_{i}).
\end{equation}

\end {itemize}

Decoding:

Receiver 2: Given a vector $z^{N}\in \mathcal{Z}^{N}$, try to find a sequence $u^{N}(\hat{t},\hat{i})$ such that
$(u^{N}(\hat{t},\hat{i}),z^{N})\in T^{N}_{UZ}(\epsilon_{3})$.
If there exist sequences with the same $\hat{t}$, put out the corresponding $\hat{t}$.
Otherwise, i.e., if no such sequence exists or multiple sequences have different message indices,
declare a decoding error.

Receiver 1: Given a vector $y^{N}\in \mathcal{Y}^{N}$, try to find a sequence $u^{N}$ such that
$(u^{N},y^{N})\in T^{N}_{UY}(\epsilon_{4})$.
If such a sequence does not exist, or there are more than one such sequence, declare a decoding error.
Denote the corresponding sequence by $u^{N}(\hat{t},\hat{i})$, put out the corresponding index $\hat{t}$.

After decoding $u^{N}(\hat{t},\hat{i})$ and $\hat{t}$, try to find a sequence $k^{N}(u^{N}(\hat{t},\hat{i}))$ such that
$(k^{N}(u^{N}(\hat{t},\hat{i})),y^{N})\in T^{N}_{KY|U}(\epsilon_{5})$.
If there exist sequences with the same index of the bin $\hat{s}$, put out the corresponding $\hat{s}$.
Otherwise, declare a decoding error.

\subsection{Proof of Achievability\label{app1.2}}

By using the above definitions, it is easy to verify that  $\lim_{N\rightarrow \infty}\frac{\log\parallel \mathcal{T}\parallel}{N}= R_{0}$ and
$\lim_{N\rightarrow \infty}\frac{\log\parallel \mathcal{S}\parallel}{N}= R_{1}$.

Then, note that the above encoding and decoding scheme is similar to the one used in \cite{St}. Hence, by similar
arguments as in \cite{St}, it is easy to show that $P_{e1}\leq \epsilon$ and $P_{e2}\leq \epsilon$, and the proof is omitted here.
It remains to show that $\lim_{N\rightarrow \infty}\Delta\geq R_{e}$, see the following.

\begin{eqnarray}\label{eapp5}
\lim_{N\rightarrow \infty}\Delta&=&\lim_{N\rightarrow \infty}\frac{1}{N}H(S|Z^{N})\nonumber\\
&\geq&\lim_{N\rightarrow \infty}\frac{1}{N}H(S|Z^{N},U^{N})\nonumber\\
&=&\lim_{N\rightarrow \infty}\frac{1}{N}(H(S,Z^{N},U^{N})-H(Z^{N},U^{N}))\nonumber\\
&=&\lim_{N\rightarrow \infty}\frac{1}{N}(H(S,Z^{N},U^{N},J,K^{N})-H(J,K^{N}|Z^{N},U^{N},S)-H(Z^{N},U^{N}))\nonumber\\
&\stackrel{(a)}=&\lim_{N\rightarrow \infty}\frac{1}{N}(H(Z^{N}|U^{N},K^{N})+H(U^{N},J,K^{N},S)-H(J,K^{N}|Z^{N},U^{N},S)-H(Z^{N},U^{N}))\nonumber\\
&\stackrel{(b)}=&\lim_{N\rightarrow \infty}\frac{1}{N}(H(Z^{N}|U^{N},K^{N})+H(U^{N},K^{N})-H(J,K^{N}|Z^{N},U^{N},S)-H(Z^{N},U^{N}))\nonumber\\
&=&\lim_{N\rightarrow \infty}\frac{1}{N}(H(K^{N}|U^{N})-H(J,K^{N}|Z^{N},U^{N},S)-I(Z^{N};K^{N}|U^{N}))\nonumber\\
&=&\lim_{N\rightarrow \infty}\frac{1}{N}(H(K^{N}|U^{N})-H(J|Z^{N},U^{N},S)-H(K^{N}|Z^{N},U^{N},S,J)-I(Z^{N};K^{N}|U^{N}))\nonumber\\
&\stackrel{(c)}\geq&\lim_{N\rightarrow \infty}\frac{1}{N}(H(K^{N}|U^{N})-\log \|\mathcal{J}\|-H(K^{N}|Z^{N},U^{N},S,J)-I(Z^{N};K^{N}|U^{N}))\nonumber\\
&\geq&\lim_{N\rightarrow \infty}\frac{1}{N}(H(K^{N}|U^{N})-H(K^{N}|U^{N},Y^{N})-\log \|\mathcal{J}\|-H(K^{N}|Z^{N},U^{N},S,J)-I(Z^{N};K^{N}|U^{N}))\nonumber\\
&=&\lim_{N\rightarrow \infty}\frac{1}{N}(I(Y^{N};K^{N}|U^{N})-\log \|\mathcal{J}\|-H(K^{N}|Z^{N},U^{N},S,J)-I(Z^{N};K^{N}|U^{N}))\nonumber\\
&\stackrel{(d)}=&\lim_{N\rightarrow \infty}\frac{1}{N}(NI(Y;K|U)-\log \|\mathcal{J}\|-H(K^{N}|Z^{N},U^{N},S,J)-NI(Z;K|U))\nonumber\\
&\stackrel{(e)}=&\lim_{N\rightarrow \infty}\frac{1}{N}(NI(Y;K|U)-N\max(I(K;V|U), I(K;Z|U))+NI(K;Z|U)-N\epsilon_{3,N}-NI(Z;K|U))\nonumber\\
&\stackrel{(f)}=&I(Y;K|U)-\max(I(K;V|U), I(K;Z|U))=R_{e},
\end{eqnarray}
where (a) is from $(S,J)\rightarrow (U^{N},K^{N})\rightarrow Z^{N}$, (b) is from $H(J,S|U^{N},K^{N})=0$,
(c) is from $H(J|Z^{N},U^{N},S)\leq H(J)\leq \log \|\mathcal{J}\|$, (d) is from that $V^{N}$, $U^{N}$, $K^{N}$ and $X^{N}$ are
i.i.d. generated random vectors, and the channels are discrete memoryless, (e) is from (\ref{eapp4x}) and (\ref{eapp4}), and
(f) is from $\epsilon_{3,N}\rightarrow 0$ as $N\rightarrow \infty$.

Thus, $\lim_{N\rightarrow \infty}\Delta\geq R_{e}$ is proved, and the proof of Theorem \ref{T1} is completed.

\section{Proof of Theorem \ref{T2}\label{appen2}}

In this section, we prove Theorem \ref{T2}: all the achievable $(R_{0}, R_{1}, R_{e})$ triples are
contained in the set $\mathcal{R}^{(no)}$.
Suppose $(R_{0}, R_{1}, R_{e})$ is achievable, i.e., for any given $\epsilon>
0$, there exists a channel encoder-decoder $(N, \Delta, P_{e1}, P_{e2})$ such that
$$\lim_{N\rightarrow \infty}\frac{\log\parallel \mathcal{T}\parallel}{N}= R_{0}, \lim_{N\rightarrow \infty}\frac{\log\parallel \mathcal{S}\parallel}{N}= R_{1},
\lim_{N\rightarrow \infty}\Delta\geq R_{e}, P_{e1}\leq \epsilon, P_{e2}\leq \epsilon.$$
Then we will show the existence of random variables $(U, K, A)\rightarrow (X, V)\rightarrow Y\rightarrow Z$ such that
\begin{equation}\label{eapp2-1}
0\leq R_{e}\leq R_{1},
\end{equation}
\begin{equation}\label{eapp2-2}
R_{0}\leq I(U;Z)-I(U;V),
\end{equation}
\begin{equation}\label{eapp2-3}
R_{1}\leq I(K;Y|U,A)-I(K;V|U,A),
\end{equation}
\begin{equation}\label{eapp2-4}
R_{0}+R_{1}\leq I(U,K,A;Y)-I(U,K,A;V),
\end{equation}
\begin{equation}\label{eapp2-5}
R_{e}\leq I(K;Y|U,A)-I(K;V|U,A)-I(K;Z|U)+I(K;V|U).
\end{equation}

Since $S$ and $T$ are independent and uniformly distributed over $\mathcal{S}$ and $\mathcal{T}$, we have $H(S)=\log\parallel \mathcal{S}\parallel$
and $H(T)=\log\parallel \mathcal{T}\parallel$.
The formulas (\ref{eapp2-2}), (\ref{eapp2-3}), (\ref{eapp2-4}) and (\ref{eapp2-5}) are proved by
Lemma \ref{L1}, see the following.

\begin{lemma}\label{L1} The random vectors $Y^{N}$, $Z^{N}$ and the random variables $S$, $T$, $V$,
$U$, $K$, $A$, $Y$, $Z$ of Theorem \ref{T2}, satisfy:
\begin{equation}\label{e307}
\frac{1}{N}H(T)\leq I(U;Z)-I(U;V)+\frac{1}{N}\delta(P_{e2}),
\end{equation}
\begin{equation}\label{e305}
\frac{1}{N}H(S)=\frac{1}{N}H(S|T)\leq I(K;Y|U,A)-I(K;V|U,A)+\frac{1}{N}\delta(P_{e1}),
\end{equation}
\begin{equation}\label{e305.1}
\frac{1}{N}H(S,T)\leq I(U,K,A;Y)-I(U,K,A;V)+\frac{1}{N}\delta(P_{e1}),
\end{equation}
\begin{eqnarray}\label{e306}
\frac{1}{N}H(S|Z^{N})&\leq& I(K;Y|U,A)-I(K;V|U,A)-I(K;Z|U)+I(K;V|U)\nonumber\\
&+&\frac{1}{N}\delta(P_{e1})+\frac{1}{N}\delta(P_{e2}),
\end{eqnarray}
where $\delta(P_{e1})=h(P_{e1})+P_{e1}\log(|\mathcal{S}\times \mathcal{T}|-1)$ and
$\delta(P_{e2})=h(P_{e2})+P_{e2}\log(|\mathcal{T}|-1)$. Note that $h(P_{e1})=-P_{e1}\log P_{e1}-(1-P_{e1})\log(1-P_{e1})$ and
$h(P_{e2})=-P_{e2}\log P_{e2}-(1-P_{e2})\log(1-P_{e2})$.
\end{lemma}

Substituting $H(S)=\log\parallel \mathcal{S}\parallel$, $H(T)=\log\parallel \mathcal{T}\parallel$, $H(S,T)=H(S)+H(T)$
and (\ref{e202}) into (\ref{e307}), (\ref{e305}), (\ref{e305.1}) and (\ref{e306}),
and using the fact that $\epsilon\rightarrow 0$,
the formulas (\ref{eapp2-2}), (\ref{eapp2-3}), (\ref{eapp2-4}) and (\ref{eapp2-5}) are obtained.
The formula (\ref{eapp2-1}) is from
$$R_{e}\leq \lim_{N\rightarrow \infty}\Delta=\lim_{N\rightarrow \infty}\frac{1}{N}H(S|Z^{N})\leq \lim_{N\rightarrow \infty}\frac{1}{N}H(S)=R_{1}.$$

It remains to prove Lemma \ref{L1}, see the following.
\begin{IEEEproof}[Proof of Lemma \ref{L1}] The formula (\ref{e307}) follows from
(\ref{e312}), (\ref{e315}) and (\ref{e331}). The formula (\ref{e305})
is from (\ref{e313}), (\ref{e320}) and (\ref{e333}).
The formula (\ref{e305.1})
is from (\ref{e313.1}), (\ref{e323.1}) and (\ref{e333.1}).
The formula (\ref{e306}) is proved by (\ref{e314}), (\ref{e320}),
(\ref{e324}), (\ref{e333}) and (\ref{e334}).

$<$Part i$>$ We begin with the left parts of the inequalities (\ref{e307}),
(\ref{e305}), (\ref{e305.1}) and (\ref{e306}), see the following.

Since $T\rightarrow Y^{N}\rightarrow Z^{N}$ is a Markov chain, for the common message $T$, we have
\begin{eqnarray}\label{e312}
\frac{1}{N}H(T)&=&\frac{1}{N}H(T|Z^{N})+\frac{1}{N}I(Z^{N};T)\nonumber\\
&\leq^{(a)}&\frac{1}{N}\delta(P_{e2})+\frac{1}{N}I(Z^{N};T).
\end{eqnarray}

For the confidential message $S$, we have
\begin{eqnarray}\label{e313}
\frac{1}{N}H(S)&=&\frac{1}{N}H(S|T)=\frac{1}{N}I(S;Y^{N}|T)+\frac{1}{N}H(S|Y^{N},T)\nonumber\\
&\leq&\frac{1}{N}I(S;Y^{N}|T)+\frac{1}{N}H(S,T|Y^{N})\nonumber\\
&\leq^{(b)}&\frac{1}{N}I(S;Y^{N}|T)+\frac{1}{N}\delta(P_{e1}).
\end{eqnarray}

For $S$ and $T$, we have
\begin{eqnarray}\label{e313.1}
\frac{1}{N}(H(S)+H(T))&=&\frac{1}{N}H(S,T)=\frac{1}{N}I(S,T;Y^{N})+\frac{1}{N}H(S,T|Y^{N})\nonumber\\
&\leq^{(c)}&\frac{1}{N}I(S,T;Y^{N})+\frac{1}{N}\delta(P_{e1}).
\end{eqnarray}

For the equivocation to the receiver 2, we have
\begin{eqnarray}\label{e314}
\frac{1}{N}H(S|Z^{N})&=&\frac{1}{N}H(S|Z^{N},T)+\frac{1}{N}I(S;T|Z^{N})\nonumber\\
&=&\frac{1}{N}H(S|T)-\frac{1}{N}I(S;Z^{N}|T)+\frac{1}{N}I(S;T|Z^{N})\nonumber\\
&=&\frac{1}{N}I(S;Y^{N}|T)+\frac{1}{N}H(S|Y^{N},T)-\frac{1}{N}I(S;Z^{N}|T)+\frac{1}{N}I(S;T|Z^{N})\nonumber\\
&\leq&\frac{1}{N}I(S;Y^{N}|T)-\frac{1}{N}I(S;Z^{N}|T)+\frac{1}{N}H(S,T|Y^{N})+\frac{1}{N}H(T|Z^{N})\nonumber\\
&\leq^{(d)}&\frac{1}{N}I(S;Y^{N}|T)-\frac{1}{N}I(S;Z^{N}|T)+\frac{1}{N}\delta(P_{e1})+\frac{1}{N}\delta(P_{e2}).
\end{eqnarray}

Note that (a), (b), (c) and (d) follow from Fano's inequality.

$<$Part ii$>$ By using chain rule, the character $I(Z^{N};T)$ in formula (\ref{e312})
can be bounded as follows,
\begin{eqnarray}\label{e315}
\frac{1}{N}I(Z^{N};T)&=&\frac{1}{N}\sum_{i=1}^{N}I(Z_{i};T|Z^{i-1})\nonumber\\
&=^{(1)}&\frac{1}{N}\sum_{i=1}^{N}(I(Z_{i};T|Z^{i-1})-I(V_{i};T|V_{i+1}^{N}))\nonumber\\
&=&\frac{1}{N}\sum_{i=1}^{N}(I(Z_{i};T,V_{i+1}^{N}|Z^{i-1})-I(Z_{i};V_{i+1}^{N}|T,Z^{i-1})-I(V_{i};T,Z^{i-1}|V_{i+1}^{N})+I(V_{i};Z^{i-1}|T,V_{i+1}^{N}))\nonumber\\
&=^{(2)}&\frac{1}{N}\sum_{i=1}^{N}(I(Z_{i};T,V_{i+1}^{N}|Z^{i-1})-I(V_{i};T,Z^{i-1}|V_{i+1}^{N}))\nonumber\\
&=^{(3)}&\frac{1}{N}\sum_{i=1}^{N}(H(Z_{i}|Z^{i-1})-H(Z_{i}|Z^{i-1},T,V_{i+1}^{N})-H(V_{i})+H(V_{i}|Z^{i-1},T,V_{i+1}^{N}))\nonumber\\
&\leq&\frac{1}{N}\sum_{i=1}^{N}(H(Z_{i})-H(Z_{i}|Z^{i-1},T,V_{i+1}^{N})-H(V_{i})+H(V_{i}|Z^{i-1},T,V_{i+1}^{N})),
\end{eqnarray}
where formula (1) follows from that $V_{i} (1\leq i\leq N)$ are i.i.d. random variables and they are independent of $T$,
formula (2) follows from that
\begin{equation}\label{e316}
\sum_{i=1}^{N}I(Z_{i};V_{i+1}^{N}|T,Z^{i-1})=\sum_{i=1}^{N}I(V_{i};Z^{i-1}|T,V_{i+1}^{N}),
\end{equation}
and formula (3) follows from that $V_{i} (1\leq i\leq N)$ are i.i.d. random variables.

$<$Part iii$>$ Using chain rule, the character $I(S;Y^{N}|T)$ in formulas (\ref{e313}) and (\ref{e314}) can be rewritten as follows,
\begin{eqnarray}\label{e320}
\frac{1}{N}I(Y^{N};S|T)&=&\frac{1}{N}\sum_{i=1}^{N}I(Y_{i};S|Y^{i-1},T)\nonumber\\
&=^{(a)}&\frac{1}{N}\sum_{i=1}^{N}(I(Y_{i};S|Y^{i-1},T)-I(V_{i};S|V_{i+1}^{N},T))\nonumber\\
&=&\frac{1}{N}\sum_{i=1}^{N}(I(Y_{i};S,V_{i+1}^{N}|Y^{i-1},T)-I(Y_{i};V_{i+1}^{N}|T,Y^{i-1},S)-I(V_{i};S,Y^{i-1}|V_{i+1}^{N},T)+\nonumber\\
&&I(V_{i};Y^{i-1}|T,S,V_{i+1}^{N}))\nonumber\\
&=^{(b)}&\frac{1}{N}\sum_{i=1}^{N}(I(Y_{i};S,V_{i+1}^{N}|Y^{i-1},T)-I(V_{i};S,Y^{i-1}|V_{i+1}^{N},T))\nonumber\\
&=&\frac{1}{N}\sum_{i=1}^{N}(H(S,V_{i+1}^{N}|Y^{i-1},T)-H(S,V_{i+1}^{N}|Y^{i-1},T,Y_{i})-H(S,Y^{i-1}|V_{i+1}^{N},T)+\nonumber\\
&&H(S,Y^{i-1}|V_{i+1}^{N},T,V_{i}))\nonumber\\
&=&\frac{1}{N}\sum_{i=1}^{N}(H(V_{i+1}^{N}|Y^{i-1},T)+H(S|Y^{i-1},T,V_{i+1}^{N})-H(V_{i+1}^{N}|Y^{i-1},T,Y_{i})-\nonumber\\
&&H(S|Y^{i-1},T,Y_{i},V_{i+1}^{N})-H(Y^{i-1}|V_{i+1}^{N},T)-H(S|V_{i+1}^{N},T,Y^{i-1})+\nonumber\\
&&H(Y^{i-1}|V_{i+1}^{N},T,V_{i})+H(S|V_{i+1}^{N},T,V_{i},Y^{i-1}))\nonumber\\
&=&\frac{1}{N}\sum_{i=1}^{N}(I(V_{i+1}^{N};Y_{i}|Y^{i-1},T)-I(Y^{i-1};V_{i}|V_{i+1}^{N},T)+\nonumber\\
&&I(S;Y_{i}|Y^{i-1},T,V_{i+1}^{N})-I(S;V_{i}|Y^{i-1},V_{i+1}^{N},T))\nonumber\\
&=^{(c)}&\frac{1}{N}\sum_{i=1}^{N}(I(S;Y_{i}|Y^{i-1},T,V_{i+1}^{N})-I(S;V_{i}|Y^{i-1},V_{i+1}^{N},T))\nonumber\\
&=^{(d)}&\frac{1}{N}\sum_{i=1}^{N}(I(S;Y_{i}|Y^{i-1},T,V_{i+1}^{N},Z^{i-1})-I(S;V_{i}|Y^{i-1},V_{i+1}^{N},T,Z^{i-1})),
\end{eqnarray}
where formula (a) follows from $V_{i} (1\leq i\leq N)$ are i.i.d. random variables and they are independent of $T$, $S$,
formula (b) follows from
\begin{equation}\label{e321}
\sum_{i=1}^{N}I(Y_{i};V_{i+1}^{N}|T,Y^{i-1},S)=\sum_{i=1}^{N}I(V_{i};Y^{i-1}|T,S,V_{i+1}^{N}),
\end{equation}
formula (c) follows from (\ref{e316}), and formula (d) is from the Markov chains
$Y_{i}\rightarrow (Y^{i-1},T,V_{i+1}^{N})\rightarrow Z^{i-1}$, $Y_{i}\rightarrow (Y^{i-1},T,V_{i+1}^{N},S)\rightarrow Z^{i-1}$,
$V_{i}\rightarrow (Y^{i-1},T,V_{i+1}^{N})\rightarrow Z^{i-1}$ and
$V_{i}\rightarrow (Y^{i-1},T,V_{i+1}^{N},S)\rightarrow Z^{i-1}$.

$<$Part iv$>$ Similar to (\ref{e315}), the character $I(S,T;Y^{N})$ in formula (\ref{e313.1}) can be rewritten as follows,
\begin{equation}\label{e323.1}
\frac{1}{N}I(S,T;Y^{N})\leq\frac{1}{N}\sum_{i=1}^{N}(H(Y_{i})-H(Y_{i}|Y^{i-1},T,S,V_{i+1}^{N})-H(V_{i})+H(V_{i}|Y^{i-1},T,S,V_{i+1}^{N})).
\end{equation}

$<$Part v$>$ Similar to (\ref{e320}), the character $I(S;Z^{N}|T)$ in formula (\ref{e314}) can be rewritten as follows,
\begin{eqnarray}\label{e324}
\frac{1}{N}I(S;Z^{N}|T)&=&\frac{1}{N}\sum_{i=1}^{N}I(Z_{i};S|Z^{i-1},T)\nonumber\\
&=&\frac{1}{N}\sum_{i=1}^{N}(I(Z_{i};S|Z^{i-1},T)-I(V_{i};S|V_{i+1}^{N},T))\nonumber\\
&=&\frac{1}{N}\sum_{i=1}^{N}(I(Z_{i};S,V_{i+1}^{N}|Z^{i-1},T)-I(Z_{i};V_{i+1}^{N}|T,Z^{i-1},S)-I(V_{i};S,Z^{i-1}|V_{i+1}^{N},T)+\nonumber\\
&&I(V_{i};Z^{i-1}|T,S,V_{i+1}^{N}))\nonumber\\
&=&\frac{1}{N}\sum_{i=1}^{N}(I(Z_{i};S,V_{i+1}^{N}|Z^{i-1},T)-I(V_{i};S,Z^{i-1}|V_{i+1}^{N},T))\nonumber\\
&=&\frac{1}{N}\sum_{i=1}^{N}(I(V_{i+1}^{N};Z_{i}|Z^{i-1},T)-I(Z^{i-1};V_{i}|V_{i+1}^{N},T)+\nonumber\\
&&I(S;Z_{i}|Z^{i-1},T,V_{i+1}^{N})-I(S;V_{i}|Z^{i-1},V_{i+1}^{N},T))\nonumber\\
&=&\frac{1}{N}\sum_{i=1}^{N}(I(S;Z_{i}|Z^{i-1},T,V_{i+1}^{N})-I(S;V_{i}|Z^{i-1},V_{i+1}^{N},T)).
\end{eqnarray}

$<$Part vi$>$ (single letter) To complete the proof, we introduce a random variable $J$, which is independent of $S$, $T$, $X^{N}$, $V^{N}$, $Y^{N}$ and $Z^{N}$.
Furthermore, $J$ is uniformly distributed over $\{1,2,...,N\}$.
Define
\begin{equation}\label{e3a}
U=(T,Z^{J-1},V^{N}_{J+1},J),
\end{equation}
\begin{equation}\label{e3b}
K=S,
\end{equation}
\begin{equation}\label{e3c}
A=(Y^{J-1},J),
\end{equation}
\begin{equation}\label{e3e}
X=X_{J}, Y=Y_{J}, Z=Z_{J}, V=V_{J}.
\end{equation}

$<$Part vii$>$ Then (\ref{e315}) can be rewritten as
\begin{eqnarray}\label{e331}
\frac{1}{N}I(T;Z^{N})&\leq&\frac{1}{N}\sum_{i=1}^{N}(H(Z_{i})-H(Z_{i}|Z^{i-1},T,V_{i+1}^{N})-H(V_{i})+H(V_{i}|Z^{i-1},T,V_{i+1}^{N}))\nonumber\\
&=&\frac{1}{N}\sum_{i=1}^{N}(H(Z_{i}|J=i)-H(Z_{i}|Z^{i-1},T,V_{i+1}^{N},J=i)-H(V_{i}|J=i)+H(V_{i}|Z^{i-1},T,V_{i+1}^{N},J=i))\nonumber\\
&=&H(Z_{J}|J)-H(Z_{J}|Z^{J-1},T,V_{J+1}^{'N},J)-H(V_{J}|J)+H(V_{J}|Z^{J-1},T,V_{J+1}^{'N},J)\nonumber\\
&=^{(a)}&H(Z_{J}|J)-H(Z_{J}|Z^{J-1},T,V_{J+1}^{N},J)-H(V_{J})+H(V_{J}|Z^{J-1},T,V_{J+1}^{N},J)\nonumber\\
&=&H(Z|J)-H(Z|U)-H(V)+H(V|U)\nonumber\\
&\leq&H(Z)-H(Z|U)-H(V)+H(V|U)\nonumber\\
&=&I(U;Z)-I(U;V),
\end{eqnarray}
where (a) follows from the fact that $V_{J}$ is independent of $J$.

\begin{IEEEproof}[Proof of $p(V_{J}=v,J=i)=p(V_{J}=v)p(J=i)$]
Since $V^{N}$ is the output of a discrete memoryless source $p_{V}(v)$, we have
\begin{equation}\label{e331.1}
p(V_{i}=v)=P(V=v).
\end{equation}

From $<$Part vi$>$, we know that the random variable $J$ is independent of $V^{N}$, and therefore,
\begin{equation}\label{e331.2}
p(V_{J}=v,J=i)=p(V_{i}=v,J=i)=p(V_{i}=v)p(J=i)=^{(1)}P(V=v)p(J=i),
\end{equation}
where (1) follows from (\ref{e331.1}).

On the other hand, the probability $p(V_{J}=v)$ can be calculated as follows,
\begin{eqnarray}\label{e331.3}
p(V_{J}=v)&=&\sum_{i=1}^{N}p(V_{J}=v,J=i)=\sum_{i=1}^{N}p(V_{i}=v,J=i)\nonumber\\
&=^{(a)}&\sum_{i=1}^{N}P(V_{i}=v)p(J=i)=^{(b)}\sum_{i=1}^{N}P(V=v)p(J=i)\nonumber\\
&=&P(V=v)\sum_{i=1}^{N}p(J=i)=P(V=v),
\end{eqnarray}
where (a) is from that $J$ is independent of $V^{N}$, the formula (b) is from (\ref{e331.1}).

By using (\ref{e331.2}) and (\ref{e331.3}), it is easy to verify that $V_{J}$ is independent of $J$,
completing the proof.
\end{IEEEproof}

Analogously, (\ref{e320}) is rewritten as follows,
\begin{eqnarray}\label{e333}
\frac{1}{N}I(Y^{N};S|T)&\leq&\frac{1}{N}\sum_{i=1}^{N}(I(S;Y_{i}|Y^{i-1},T,V_{i+1}^{N},Z^{i-1})-I(S;V_{i}|Y^{i-1},V_{i+1}^{N},T,Z^{i-1}))\nonumber\\
&=&\frac{1}{N}\sum_{i=1}^{N}(I(S;Y_{i}|Y^{i-1},T,V_{i+1}^{N},Z^{i-1},J=i)-I(S;V_{i}|Y^{i-1},V_{i+1}^{N},T,Z^{i-1},J=i))\nonumber\\
&=&I(S;Y_{J}|Y^{J-1},T,V_{J+1}^{N},Z^{J-1},J)-I(S;V_{J}|Y^{J-1},T,V_{J+1}^{N},Z^{J-1},J)\nonumber\\
&=^{(a)}&I(K;Y|U,A)-I(K;V|U,A),
\end{eqnarray}
where (a) follows from (\ref{e3a}) ,(\ref{e3b}), (\ref{e3c}) and (\ref{e3e}).

Similarly, (\ref{e323.1}) is rewritten as follows,
\begin{equation}\label{e333.1}
\frac{1}{N}I(S,T;Y^{N})\leq I(U,K,A;Y)-I(U,K,A;V),
\end{equation}
and (\ref{e324}) can be rewritten as follows,
\begin{equation}\label{e334}
\frac{1}{N}I(S;Z^{N}|T)=I(K;Z|U)-I(K;V|U).
\end{equation}

Substituting (\ref{e331}), (\ref{e333}), (\ref{e333.1}), (\ref{e334}) into
(\ref{e312}), (\ref{e313}), (\ref{e313.1}) and (\ref{e314}), Lemma \ref{L1}
is proved.
\end{IEEEproof}
The proof of  Theorem \ref{T2} is completed.

\section{Size Constraint of The Random Variables in Theorem \ref{T1}}\label{appen3}

By using the support lemma (see \cite{Cs}, p.310), it suffices to show that the random variables
$U$ and $K$ can be replaced by new ones, preserving the Markovity
$(U,K)\rightarrow (X, V)\rightarrow Y\rightarrow Z$ and the
mutual information $I(U;Z)$, $I(U;V)$, $I(K;Y|U)$, $I(K;V|U)$, $I(K;Z|U)$,
and furthermore,
the ranges of the new $U$ and $K$ satisfy:
$\|\mathcal{U}\|\|\leq \|\mathcal{X}\|\|\mathcal{V}\|+2$,
$\|\mathcal{K}\|\leq (\|\mathcal{X}\|\|\mathcal{V}\|+2)^{2}$.
The proof is in the reminder of this section.

Let
\begin{equation}\label{e601.1}
\bar{p}=p_{XV}(x,v).
\end{equation}
Define the following continuous scalar functions of $\bar{p}$ :
$$f_{XV}(\bar{p})=p_{XV}(x,v), f_{Y}(\bar{p})=H(Y), f_{Z}(\bar{p})=H(Z), f_{V}(\bar{p})=H(V).$$
Since there are $\|\mathcal{X}\|\|\mathcal{V}\|-1$ functions of
$f_{XV}(\bar{p})$, the total number of the continuous scalar
functions of $\bar{p}$ is $\|\mathcal{X}\|\|\mathcal{V}\|$+2.

Let $\bar{p}_{XV|U}=Pr\{X=x,V=v|U=u\}$. With these distributions
$\bar{p}_{XV|U}=Pr\{X=x,V=v|U=u\}$, we have
\begin{equation}\label{e602.1}
p_{XV}(x,v)=\sum_{u\in \mathcal{U}}p(U=u)f_{XV}(\bar{p}_{XV|U}),
\end{equation}
\begin{equation}\label{e603.1}
I(U;Z)=f_{Z}(\bar{p})-\sum_{u\in
\mathcal{U}}p(U=u)f_{Z}(\bar{p}_{XV|U}),
\end{equation}
\begin{equation}\label{e604.1}
I(U;V)=f_{V}(\bar{p})-\sum_{u\in
\mathcal{U}}p(U=u)f_{V}(\bar{p}_{XV|U}),
\end{equation}
\begin{equation}\label{e605.1}
H(Y|U)=\sum_{u\in
\mathcal{U}}p(U=u)f_{Y}(\bar{p}_{XV|U}),
\end{equation}

According to the support lemma (\cite{Cs}, p.310), the random
variable $U$ can be replaced by new ones such that the new $U$ takes
at most $\|\mathcal{X}\|\|\mathcal{V}\|+2$ different values and the
expressions (\ref{e602.1}), (\ref{e603.1}), (\ref{e604.1}) and (\ref{e605.1}) are preserved.

Once the alphabet of $U$ is fixed, we apply similar arguments to bound the alphabet of $K$, see the following.
Define $\|\mathcal{X}\|\|\mathcal{V}\|+2$ continuous scalar functions of $\bar{p}_{XV}$ :
$$f_{XV}(\bar{p}_{XV})=p_{XV}(x,v), f_{Y}(\bar{p}_{XV})=H(Y), f_{Z}(\bar{p}_{XV})=H(Z), f_{V}(\bar{p}_{XV})=H(V),$$
where of the functions $f_{XV}(\bar{p}_{XV})$, only $\|\mathcal{X}\|\|\mathcal{V}\|-1$ are to be considered.

For every fixed $u$, let $\bar{p}_{XV|K}=Pr\{X=x,V=v|K=k\}$. With these distributions $\bar{p}_{XV|K}$, we have
\begin{equation}\label{e607.1}
Pr\{X=x,V=v|U=u\}=\sum_{k\in \mathcal{K}}Pr\{K=k|U=u\}f_{XV}(\bar{p}_{XV|K}),
\end{equation}
\begin{equation}\label{e608.1}
I(K;Z|U)=H(Z|U=u)-\sum_{k\in \mathcal{K}}f_{Z}(\bar{p}_{XV|K})Pr\{K=k|U=u\},
\end{equation}
\begin{equation}\label{e609.1}
I(K;V|U)=H(V|U=u)-\sum_{k\in \mathcal{K}}f_{V}(\bar{p}_{XV|K})Pr\{K=k|U=u\}.
\end{equation}
\begin{equation}\label{e610.1}
I(K;Y|U)=H(Y|U=u)-\sum_{k\in \mathcal{K}}f_{Y}(\bar{p}_{XV|K})Pr\{K=k|U=u\}.
\end{equation}

By the support lemma (\cite{Cs}, p.310), for every fixed $u$, the size of the alphabet of the random variable $K$ can not be larger than
$\|\mathcal{X}\|\|\mathcal{V}\|+2$, and therefore,
$\|\mathcal{K}\|\leq (\|\mathcal{X}\|\|\mathcal{V}\|+2)^{2}$ is proved.

\section{Proof of Theorem \ref{T3}\label{appen4}}

In this section, we will show that any triple
$(R_{0},R_{1},R_{e})\in \mathcal{R}^{ci}$ is achievable.
Superposition coding and Wyner's random binning techniques are used in the construction of the code-books.

Now the
remainder of this section is organized as follows.
The code construction is
in Subsection \ref{app3.1}. The proof of achievability is given in Subsection \ref{app3.2}.

\subsection{Code Construction\label{app3.1}}

Since $R_{e}\leq I(K;Y|U)-I(K;Z|U)$ and $R_{e}\leq R_{1}\leq I(K;Y|U)$, it is sufficient to show that
the triple $(R_{0},R_{1},R_{e}=I(K;Y|U)-I(K;Z|U))$ is achievable, and note that this implies that
$R_{1}\geq R_{e}=I(K;Y|U)-I(K;Z|U)$.

Given a triple $(R_{0},R_{1},R_{e})$, choose a joint probability mass function $p_{U,K,V,X,Y,Z}(u,k,v,x,y,z)$
such that
$$0\leq R_{e}\leq R_{1},$$
$$R_{0}\leq I(U;Z),$$
$$R_{1}\leq I(K;Y|U),$$
$$R_{e}=I(K;Y|U)-I(K;Z|U).$$

The confidential message set $\mathcal{S}$ and the common message set $\mathcal{T}$ satisfy the following conditions:
\begin{equation}\label{eapp1a}
\lim_{N\rightarrow \infty}\frac{1}{N}\log\parallel \mathcal{S}\parallel=R_{1}=I(K;Y|U)-\gamma_{1},
\end{equation}
\begin{equation}\label{eapp2a}
\lim_{N\rightarrow \infty}\frac{1}{N}\log\parallel \mathcal{T}\parallel= R_{0}=I(U;Z)-\gamma,
\end{equation}
where $\gamma$ and $\gamma_{1}$ are fixed numbers and $\gamma\geq 0$,
\begin{equation}\label{eapp2.5a}
0\leq \gamma_{1}\leq^{(a)} I(K;Z|U).
\end{equation}
Note that (a) is from $R_{1}\geq R_{e}=I(K;Y|U)-I(K;Z|U)$ and (\ref{eapp1a}).
Let $\mathcal{S}=\{1,2,...,2^{NR_{1}}\}$ and $\mathcal{T}=\{1,2,...,2^{NR_{0}}\}$.

Code-book generation:

\begin{itemize}

\item \textbf{(Construction of $U^{N}$)}

For a given common message $t$ ($t\in \mathcal{T}$), generate a corresponding sequence
$u^{N}(t)$ i.i.d. according to the probability mass function
$p_{U}(u)$.

\item \textbf{(Construction of $K^{N}$)}
Classical superposition coding and Wyner's random binning technique \cite{Wy} are used in the construction of $K^{N}$, see Figure \ref{f5}.

\begin{figure}[htb]
\centering
\includegraphics[scale=0.6]{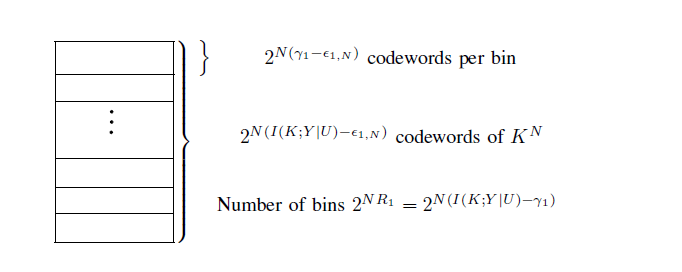}
\caption{Code-book construction for $K^{N}$ in Theorem \ref{T3}}
\label{f5}
\end{figure}

For the transmitted sequence $u^{N}(t)$, generate $2^{N(I(K;Y|U)-\epsilon_{1,N})}$
($\epsilon_{1,N}\rightarrow 0$ as $N\rightarrow \infty$) i.i.d. sequences $k^{N}$,
according to the probability mass function $p_{K|U}(k_{i}|u_{i}(t))$.
Distribute these sequences at random into $2^{NR_{1}}=2^{N(I(K;Y|U)-\gamma_{1})}$ bins such that
each bin contains $2^{N(\gamma_{1}-\epsilon_{1,N})}$ sequences. Index each bin by $i\in \{1,2,...,2^{NR_{1}}\}$.

Here note that the number of the sequences in every bin is upper bounded as follows.
\begin{equation}\label{eapp3a}
\gamma_{1}-\epsilon_{1,N}\leq^{(a)}I(K;Z|U)-\epsilon_{1,N},
\end{equation}
where (a) is from (\ref{eapp2.5a}). This implies that
\begin{equation}\label{eapp4a}
\lim_{N\rightarrow \infty}H(K^{N}|U^{N},S,Z^{N})=0.
\end{equation}

For a given confidential message $s$ ($s\in \mathcal{S}$), randomly choose
a sequence $k^{N}(u^{N}(t))$ in bin $s$ for transmission.

\item \textbf{(Construction of $X^{N}$)}
The $x^{N}$ is generated
according to a new discrete memoryless channel (DMC) with inputs $k^{N}$, $u^{N}$, $v^{N}$,
and output $x^{N}$. The transition probability of this new DMC is $p_{X|U,K,V}(x|u,k,v)$.
The probability $p_{X^{N}|U^{N},K^{N},V^{N}}(x^{N}|u^{N},k^{N},v^{N})$ is calculated as follows.
\begin{equation}\label{eapp4.5}
p_{X^{N}|U^{N},K^{N},V^{N}}(x^{N}|u^{N},k^{N},v^{N})=
\prod_{i=1}^{N}p_{X|U,K,V}(x_{i}|u_{i},k_{i},v_{i}).
\end{equation}

\end {itemize}

Decoding:

Receiver 2: Given a vector $z^{N}\in \mathcal{Z}^{N}$, try to find a sequence $u^{N}(\hat{t})$ such that
$(u^{N}(\hat{t}),z^{N})\in T^{N}_{UZ}(\epsilon)$.
If there exists a unique sequence, put out the corresponding $\hat{t}$.
Otherwise, declare a decoding error.

Receiver 1: Given a vector $y^{N}\in \mathcal{Y}^{N}$, try to find a sequence $u^{N}$ such that
$(u^{N},y^{N})\in T^{N}_{UY}(\epsilon)$.
If such a sequence does not exist, or there are more than one such sequence, declare a decoding error.
Denote the corresponding sequence by $u^{N}(\hat{t})$, put out the corresponding index $\hat{t}$.

After decoding $u^{N}(\hat{t})$ and $\hat{t}$, try to find a sequence $k^{N}(u^{N}(\hat{t}))$ such that
$(k^{N}(u^{N}(\hat{t})),y^{N})\in T^{N}_{KY|U}(\epsilon)$.
If there exist sequences with the same index of the bin $\hat{s}$, put out the corresponding $\hat{s}$.
Otherwise, declare a decoding error.

\subsection{Proof of Achievability\label{app3.2}}

By using the above definitions, it is easy to verify that  $\lim_{N\rightarrow \infty}\frac{\log\parallel \mathcal{T}\parallel}{N}= R_{0}$ and
$\lim_{N\rightarrow \infty}\frac{\log\parallel \mathcal{S}\parallel}{N}= R_{1}$.

Then, note that the above encoding and decoding scheme is similar to the one used in \cite{St}. Hence, by similar
arguments as in \cite{St}, it is easy to show that $P_{e1}\leq \epsilon$ and $P_{e2}\leq \epsilon$, and the proof is omitted here.
It remains to show that $\lim_{N\rightarrow \infty}\Delta\geq R_{e}$, see the following.

\begin{eqnarray}\label{eapp5a}
\lim_{N\rightarrow \infty}\Delta&=&\lim_{N\rightarrow \infty}\frac{1}{N}H(S|Z^{N})\nonumber\\
&\geq&\lim_{N\rightarrow \infty}\frac{1}{N}H(S|Z^{N},U^{N})\nonumber\\
&=&\lim_{N\rightarrow \infty}\frac{1}{N}(H(S,Z^{N},U^{N})-H(Z^{N},U^{N}))\nonumber\\
&=&\lim_{N\rightarrow \infty}\frac{1}{N}(H(S,Z^{N},U^{N},K^{N})-H(K^{N}|Z^{N},U^{N},S)-H(Z^{N},U^{N}))\nonumber\\
&\stackrel{(a)}=&\lim_{N\rightarrow \infty}\frac{1}{N}(H(Z^{N}|U^{N},K^{N})+H(U^{N},K^{N},S)-H(K^{N}|Z^{N},U^{N},S)-H(Z^{N},U^{N}))\nonumber\\
&\stackrel{(b)}=&\lim_{N\rightarrow \infty}\frac{1}{N}(H(Z^{N}|U^{N},K^{N})+H(U^{N},K^{N})-H(K^{N}|Z^{N},U^{N},S)-H(Z^{N},U^{N}))\nonumber\\
&=&\lim_{N\rightarrow \infty}\frac{1}{N}(H(K^{N}|U^{N})-H(K^{N}|Z^{N},U^{N},S)-I(Z^{N};K^{N}|U^{N}))\nonumber\\
&\geq&\lim_{N\rightarrow \infty}\frac{1}{N}(H(K^{N}|U^{N})-H(K^{N}|U^{N},Y^{N})-H(K^{N}|Z^{N},U^{N},S)-I(Z^{N};K^{N}|U^{N}))\nonumber\\
&=&\lim_{N\rightarrow \infty}\frac{1}{N}(I(Y^{N};K^{N}|U^{N})-H(K^{N}|Z^{N},U^{N},S)-I(Z^{N};K^{N}|U^{N}))\nonumber\\
&\stackrel{(c)}=&\lim_{N\rightarrow \infty}\frac{1}{N}(NI(Y;K|U)-H(K^{N}|Z^{N},U^{N},S)-NI(K;Z|U))\nonumber\\
&\stackrel{(d)}=&\lim_{N\rightarrow \infty}\frac{1}{N}(NI(Y;K|U)-NI(K;Z|U))\nonumber\\
&=&I(Y;K|U)-I(K;Z|U)=R_{e},
\end{eqnarray}
where (a) is from $S\rightarrow (U^{N},K^{N})\rightarrow Z^{N}$, (b) is from $H(S|U^{N},K^{N})=0$,
(c) is from that $V^{N}$, $U^{N}$, $K^{N}$ and $X^{N}$ are
i.i.d. generated random vectors, and the channels are discrete memoryless, and (d) is from (\ref{eapp4a}).

Thus, $\lim_{N\rightarrow \infty}\Delta\geq R_{e}$ is proved, and the proof of Theorem \ref{T3} is completed.

\section{Proof of Theorem \ref{T4}\label{appen5}}

In this section, we prove Theorem \ref{T4}: all the achievable $(R_{0}, R_{1}, R_{e})$ triples are
contained in the set $\mathcal{R}^{(co)}$.
Suppose $(R_{0}, R_{1}, R_{e})$ is achievable, i.e., for any given $\epsilon>
0$, there exists a channel encoder-decoder $(N, \Delta, P_{e1}, P_{e2})$ such that
$$\lim_{N\rightarrow \infty}\frac{\log\parallel \mathcal{T}\parallel}{N}= R_{0}, \lim_{N\rightarrow \infty}\frac{\log\parallel \mathcal{S}\parallel}{N}= R_{1},
\lim_{N\rightarrow \infty}\Delta\geq R_{e}, P_{e1}\leq \epsilon, P_{e2}\leq \epsilon.$$
Then we will show the existence of random variables $(U, K, A)\rightarrow (X, V)\rightarrow Y\rightarrow Z$ such that
\begin{equation}\label{e335}
0\leq R_{e}\leq R_{1},
\end{equation}
\begin{equation}\label{e336}
R_{0}\leq I(U;Z),
\end{equation}
\begin{equation}\label{e337}
R_{1}\leq I(K;Y|U),
\end{equation}
\begin{equation}\label{e338}
R_{e}\leq I(K;Y|U)-I(K;Z|A).
\end{equation}

The formula (\ref{e335}) is from
$$R_{e}\leq \lim_{N\rightarrow \infty}\Delta=\lim_{N\rightarrow \infty}\frac{1}{N}H(S|Z^{N})\leq \lim_{N\rightarrow \infty}\frac{1}{N}H(S)=R_{1}.$$

Since the model of Figure \ref{f2} with causal side information is a special case of the model of Figure \ref{f2} with noncausal side information,
the formulas (\ref{e336}), (\ref{e337}) and (\ref{e338}) are obtained from (\ref{e339}), (\ref{e340})
 and (\ref{e341}), respectively, see the following.

\begin{IEEEproof}[Proof of (\ref{e336})]
The parameter $R_{0}$ of (\ref{e336}) can be written as follows,
\begin{eqnarray}\label{e339}
R_{0}&=&\lim_{N\rightarrow \infty}\frac{H(T)}{N}\nonumber\\
&\leq^{(a)}&\lim_{N\rightarrow \infty}(\frac{1}{N}\delta(P_{e2})+\frac{1}{N}\sum_{i=1}^{N}(H(Z_{i})-H(Z_{i}|Z^{i-1},T,V_{i+1}^{N})))\nonumber\\
&\leq&\lim_{N\rightarrow \infty}(\frac{1}{N}\delta(P_{e2})+\frac{1}{N}\sum_{i=1}^{N}(H(Z_{i})-H(Z_{i}|Y^{i-1},Z^{i-1},T,V_{i+1}^{N})))\nonumber\\
&\leq^{(b)}&\lim_{N\rightarrow \infty}(\frac{1}{N}\delta(\epsilon)+H(Z)-H(Z|U))\nonumber\\
&=^{(c)}&I(U;Z),
\end{eqnarray}
where (a) follows from (\ref{e312}) and (\ref{e315}), and $V_{i}$ is independent of $(Z^{i-1},T,V_{i+1}^{N})$,
the formula (b) is from the definitions $U\triangleq (T,V_{J+1}^{N},Y^{J-1},Z^{J-1},J)$, $Z\triangleq Z_{J}$,
and the formula (c) follows from $\epsilon\rightarrow 0$.
Thus, the proof of (\ref{e336}) is completed.
\end{IEEEproof}

\begin{IEEEproof}[Proof of (\ref{e337})]
The parameter $R_{1}$ of (\ref{e337}) can be written follows,
\begin{eqnarray}\label{e340}
R_{1}&=&\lim_{N\rightarrow \infty}\frac{H(S)}{N}\nonumber\\
&\leq^{(1)}&\lim_{N\rightarrow \infty}(\frac{1}{N}(I(S;Y^{N}|T)+\frac{1}{N}\delta(P_{e1}))\nonumber\\
&\leq&\lim_{N\rightarrow \infty}(\frac{1}{N}I(S;Y^{N}|T)+\frac{1}{N}\delta(\epsilon))\nonumber\\
&\leq^{(2)}&\lim_{N\rightarrow \infty}(\frac{1}{N}\sum_{i=1}^{N}I(S;Y_{i}|T,V_{i+1}^{N},Y^{i-1},Z^{i-1})+\frac{1}{N}\delta(\epsilon))\nonumber\\
&\leq^{(3)}&I(K;Y|U),
\end{eqnarray}
where (1) follows from (\ref{e313}), the formula (2) is from (\ref{e320}) and the fact that $I(S;V_{i}|T,V_{i+1}^{N},Y^{i-1},Z^{i-1})=0$,
the formula (3) is from the definitions $U\triangleq (T,V_{J+1}^{N},Y^{J-1},Z^{J-1},J)$, $K\triangleq S$, $Y\triangleq Y_{J}$, ($J$
is independent of $S$, $T$, $X^{N}$, $V^{N}$, $Y^{N}$ and $Z^{N}$,
and $J$ is uniformly distributed over $\{1,2,...,N\}$)
and the fact that $\epsilon\rightarrow 0$.
Thus, the formula (\ref{e337}) is proved.
\end{IEEEproof}

\begin{IEEEproof}[Proof of (\ref{e338})]
The parameter $R_{e}$ of (\ref{e338}) satisfies
\begin{eqnarray}\label{e341}
R_{e}&=&\lim_{N\rightarrow \infty}\frac{H(S|Z^{N})}{N}\nonumber\\
&\leq^{(a)}&\lim_{N\rightarrow \infty}(\frac{1}{N}I(S;Y^{N}|T)-\frac{1}{N}I(S;Z^{N}|T)+\frac{1}{N}\delta(P_{e1})+\frac{1}{N}\delta(P_{e2}))\nonumber\\
&\leq^{(b)}&\lim_{N\rightarrow \infty}(\frac{1}{N}\sum_{i=1}^{N}(I(S;Y_{i}|T,V_{i+1}^{N},Y^{i-1},Z^{i-1})-\nonumber\\
&&I(S;Z_{i}|Z^{i-1},T,V_{i+1}^{N}))+\frac{1}{N}\delta(P_{e1})+\frac{1}{N}\delta(P_{e2}))\nonumber\\
&=^{(c)}&I(K;Y|U)-I(K;Z|A),
\end{eqnarray}
where (a) follows from (\ref{e314}), the formula (b) is from (\ref{e320}) and (\ref{e324}), and the fact that
$V_{i}$ is independent of $(Z^{i-1},Y^{i-1},T,S,V_{i+1}^{N})$,
the formula (c) is from the definitions $U\triangleq (T,V_{J+1}^{N},Y^{J-1},Z^{J-1},J)$, $K\triangleq S$, $A\triangleq (T,V_{J+1}^{N},Z^{J-1},J)$,
$Y\triangleq Y_{J}$, $Z\triangleq Z_{J}$, and the fact that $P_{e1}, P_{e2}\leq \epsilon\rightarrow 0$.
Thus, the proof of (\ref{e338}) is completed.
\end{IEEEproof}

The proof of Theorem \ref{T4} is completed.

\section{Proof of Theorem \ref{T5}\label{appen6}}

In this section, we will show that any triple
$(R_{0},R_{1},R_{e})\in \mathcal{R}^{nfi}$ is achievable.
Superposition coding, Gel'fand-Pinsker's binning, block Markov coding and Ahlswede-Cai's secret key on feedback \cite{AC}
are used in the construction of the code-books. In addition, the encoding and decoding scheme for Theorem \ref{T5}
can be also viewed as a combination of Steinberg's method \cite{St}, block Markov coding and Ahlswede-Cai's secret key on feedback \cite{AC}.

Now the
remainder of this section is organized as follows.
The code construction is
in Subsection \ref{app6.1}. The proof of achievability is given in Subsection \ref{app6.2}.

\subsection{Code Construction\label{app6.1}}

Given a triple $(R_{0},R_{1},R_{e})$, choose a joint probability mass function $p_{U,K,V,X,Y,Z}(u,k,v,x,y,z)$
such that
$$0\leq R_{e}\leq R_{1},$$
$$R_{0}\leq I(U;Z)-I(U;V),$$
$$R_{1}\leq I(K;Y|U)-I(K;V|U),$$
$$R_{e}\leq H(Y|Z).$$

The confidential message set $\mathcal{S}$ and the common message set $\mathcal{T}$ satisfy the following conditions:
\begin{equation}\label{eapp1aa}
\lim_{N\rightarrow \infty}\frac{1}{N}\log\parallel \mathcal{S}\parallel=R_{1}=I(K;Y|U)-I(K;V|U)-\gamma_{1},
\end{equation}
\begin{equation}\label{eapp2aa}
\lim_{N\rightarrow \infty}\frac{1}{N}\log\parallel \mathcal{T}\parallel= R_{0}=I(U;Z)-I(U;V)-\gamma,
\end{equation}
where $\gamma$ and $\gamma_{1}$ are fixed positive real numbers.
Let $\mathcal{S}=\{1,2,...,2^{NR_{1}}\}$ and $\mathcal{T}=\{1,2,...,2^{NR_{0}}\}$.

We use the block Markov coding method. The random vectors $U^{N}$, $K^{N}$, $V^{N}$, $X^{N}$, $Y^{N}$
and $Z^{N}$ consist of $n$ blocks of length $N$.
The common message for $n$ blocks is $T^{n}\triangleq (T_{1},...,T_{n})$,
which is composed of $n$ i.i.d. random variables uniformly distributed over $\mathcal{T}$.
The confidential message for $n$ blocks is $S^{n}\triangleq (S_{2}, S_{3},...,S_{n})$,
where $S_{i}$ ($2\leq i\leq n$) are i.i.d. random variables uniformly distributed over $\mathcal{S}$.
Note that in the first block, there is no $S_{1}$.

Let $\widetilde{Z}_{i}$ ($1\leq i\leq n$) be the output of channel 2 for block $i$, $Z^{n}=(\widetilde{Z}_{1},...,\widetilde{Z}_{n})$,
$Z^{\overline{j}}=(\widetilde{Z}_{1},...,\widetilde{Z}_{j-1},\widetilde{Z}_{j+1},...,\widetilde{Z}_{n})$ $(1\leq j\leq n)$.
Similarly,  $Y^{n}=(\widetilde{Y}_{1},...,\widetilde{Y}_{n})$, and
$\widetilde{Y}_{i}$ ($1\leq i\leq n$) is the output of channel 1 for block $i$.
The specific values of the above random vectors are denoted by lower case letters.

Code-book generation:

\begin{itemize}

\item \textbf{(Construction of $U^{N}$)}

The construction of $U^{N}$ for each block is exactly the same as that in the proof of Theorem \ref{T1}, see the following.

For the $i$-th block ($1\leq i\leq n$),
generate $2^{N(I(U;Z)-\epsilon_{1,N})}$ ($\epsilon_{1,N}\rightarrow 0$ as $N\rightarrow \infty$)
i.i.d. sequences $u^{N}$, according to the probability mass function
$p_{U}(u)$. Distribute these sequences at random into $2^{NR_{0}}=2^{N(I(U;Z)-I(U;V)-\gamma)}$ bins such that
each bin contains $2^{N(I(U;V)+\gamma-\epsilon_{1,N})}$ sequences. Index each bin by $t_{i}\in \{1,2,...,2^{NR_{0}}\}$.

For a given common message $t_{i}$ ($t_{i}\in \mathcal{T}$) and side information $v^{N}$, try to find
a sequence in bin $t_{i}$ \\ $\{u^{N}(t_{i},1),u^{N}(t_{i},2),...,u^{N}(t_{i},2^{N(I(U;V)+\gamma-\epsilon_{1,N})})\}$
that is jointly typical with $v^{N}$, say $u^{N}(t_{i},j^{*})$, i.e., \\ $(u^{N}(t_{i},j^{*}),v^{N})\in T^{N}_{UV}(\epsilon_{1})$.
If multiple such sequences in bin $t_{i}$
exist, choose the one with the smallest $j^{*}$. If no such $j^{*}$ exists, then declare an encoding error.

\item \textbf{(Construction of $K^{N}$)}

Superposition coding, Gel'fand-Pinsker's binning, block Markov coding and Ahlswede-Cai's secret key on feedback \cite{AC}
are used in the construction of $K^{N}$, see Figure \ref{f6}.

For the transmitted sequence $u^{N}(t_{i},j^{*})$, generate $2^{N(I(K;Y|U)-\epsilon_{2,N})}$
($\epsilon_{2,N}\rightarrow 0$ as $N\rightarrow \infty$) i.i.d. sequences $k^{N}$,
according to the probability mass function $p_{K|U}(k_{i}|u_{i})$.
Distribute these sequences at random into $2^{NR_{1}}=2^{N(I(K;Y|U)-I(K;V|U)-\gamma_{1})}$ bins such that
each bin contains $2^{N(I(K;V|U)+\gamma_{1}-\epsilon_{2,N})}$ sequences. Index each bin by $l\in \{1,2,...,2^{NR_{1}}\}$.

In the first block, for a given side information $v^{N}$, try to find a  $k^{N}(u^{N}(t_{1},j^{*}))$ generated by $u^{N}(t_{1},j^{*})$
such that
$(k^{N}(u^{N}(t_{1},j^{*})),v^{N})\in T^{N}_{KV|U}(\epsilon)$. If multiple such sequences exist,
randomly choose one for transmission. If no such sequence exists, declare an encoding error.

For the $i$-th block ($2\leq i\leq n$), firstly we generate a mapping
$g_{f}: \mathcal{Y}^{N}\rightarrow \mathcal{S}$
(note that $\|\mathcal{Y}\|^{N}\geq \parallel\mathcal{S}\parallel$). Define a random variable $K_{i}^{*}=g_{f}(\widetilde{Y}_{i-1})$ ($2\leq i\leq n$),
which is uniformly
distributed over $\mathcal{S}$, and $K_{i}^{*}$ is independent of $S_{i}$.
Reveal the mapping $g_{f}$ to both receivers and the transmitter.

Then, when the transmitter receives the output $\widetilde{y}_{i-1}$ of the $i$-1-th block, he computes
$k_{i}^{*}=g_{f}(\widetilde{y}_{i-1})\in \mathcal{S}$.
For a given $s_{i}$, the transmitter chooses a sequence $k^{N}(u^{N}(t_{i},j^{*}))$ from the bin $s_{i}\oplus k_{i}^{*}$
(where $\oplus$ is the modulo addition over $\mathcal{S}$) such that
$(k^{N}(u^{N}(t_{i},j^{*})),v^{N})\in T^{N}_{KV|U}(\epsilon)$.
If multiple such sequences in bin $s_{i}\oplus k_{i}^{*}$
exist, choose the one with the smallest index in the bin. If no such sequence exists, declare an encoding error.

\begin{figure}[htb]
\centering
\includegraphics[scale=0.6]{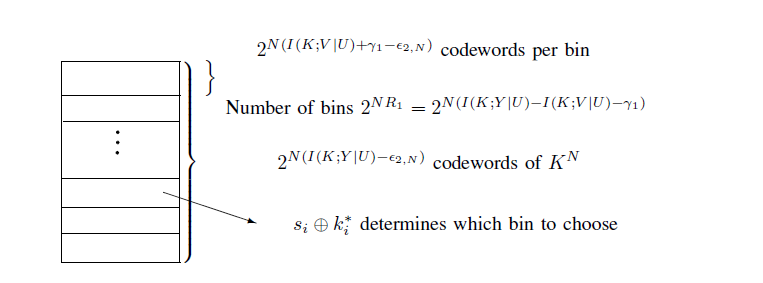}
\caption{Code-book construction for $K^{N}$ in Theorem \ref{T5}, where $s_{i}$ and $k_{i}^{*}$ are the confidential message
and the secret key for block $i$, respectively.}
\label{f6}
\end{figure}

\item \textbf{(Construction of $X^{N}$)}

For each block, the $x^{N}$ is generated
according to a new discrete memoryless channel (DMC) with inputs $k^{N}$, $u^{N}$, $v^{N}$,
and output $x^{N}$. The transition probability of this new DMC is $p_{X|U,K,V}(x|u,k,v)$.
The probability $p_{X^{N}|U^{N},K^{N},V^{N}}(x^{N}|u^{N},k^{N},v^{N})$ is calculated as follows.
\begin{equation}\label{eapp4.5xx}
p_{X^{N}|U^{N},K^{N},V^{N}}(x^{N}|u^{N},k^{N},v^{N})=
\prod_{i=1}^{N}p_{X|U,K,V}(x_{i}|u_{i},k_{i},v_{i}).
\end{equation}

\end {itemize}

Decoding:

Receiver 2: For each block, given a vector $z^{N}\in \mathcal{Z}^{N}$, try to find a sequence $u^{N}(\hat{t},\hat{i})$ such that
$(u^{N}(\hat{t},\hat{i}),z^{N})\in T^{N}_{UZ}(\epsilon_{3})$.
If there exist sequences with the same $\hat{t}$, put out the corresponding $\hat{t}$.
Otherwise, i.e., if no such sequence exists or multiple sequences have different message indices,
declare a decoding error.

Receiver 1: For the first block,  given a vector $\widetilde{y}_{1}\in \mathcal{Y}^{N}$, try to find a sequence $u^{N}$ such that
$(u^{N},\widetilde{y}_{1})\in T^{N}_{UY}(\epsilon_{4})$.
If such a sequence does not exist, or there are more than one such sequence, declare a decoding error.
Denote the corresponding sequence by $u^{N}(\hat{t}_{1},\hat{j})$, and put out the corresponding index $\hat{t}_{1}$.

For block $i$ ($2\leq i\leq n$), given a vector $\widetilde{y}_{i}\in \mathcal{Y}^{N}$, try to find a sequence $u^{N}$ such that
$(u^{N},\widetilde{y}_{i})\in T^{N}_{UY}(\epsilon_{4})$.
If such a sequence does not exist, or there are more than one such sequence, declare a decoding error.
Denote the corresponding sequence by $u^{N}(\hat{t}_{i},\hat{j})$, put out the corresponding index $\hat{t}_{i}$.

After decoding $u^{N}(\hat{t}_{i},\hat{j})$ and $\hat{t}_{i}$, try to find a sequence $k^{N}(u^{N}(\hat{t}_{i},\hat{j}))$ such that
$(k^{N}(u^{N}(\hat{t}_{i},\hat{j})),\widetilde{y}_{i})\in T^{N}_{KY|U}(\epsilon_{5})$.
If there exist sequences with the same index of the bin $\hat{s}_{i}\oplus k_{i}^{*}$, put out the corresponding $\hat{s}_{i}\oplus k_{i}^{*}$.
Otherwise, declare a decoding error. Finally, since receiver 1 knows $k_{i}^{*}$ ($k_{i}^{*}=g_{f}(\widetilde{y}_{i-1})$),
put out the corresponding $\hat{s}_{i}$ from $\hat{s}_{i}\oplus k_{i}^{*}$.

\subsection{Proof of Achievability\label{app6.2}}

The rate of the confidential message $S^{n}$ is defined as
\begin{equation}\label{e4xx.13}
R_{1}^{*}\triangleq \lim_{N\rightarrow \infty}\lim_{n\rightarrow \infty}\frac{H(S^{n})}{nN},
\end{equation}
and it satisfies
\begin{eqnarray}\label{e4xx.14}
R_{1}^{*}&=&\lim_{N\rightarrow \infty}\lim_{n\rightarrow \infty}\frac{H(S^{n})}{nN}\nonumber\\
&=&\lim_{N\rightarrow \infty}\lim_{n\rightarrow \infty}\frac{\sum_{i=2}^{n}H(S_{i})}{nN}\nonumber\\
&=&\lim_{N\rightarrow \infty}\lim_{n\rightarrow \infty}\frac{(n-1)H(S)}{nN}\nonumber\\
&=&R_{1}.
\end{eqnarray}

Similarly, the rate of the common messages $T^{n}$ satisfies
\begin{eqnarray}\label{e4xx.15}
R_{0}^{*}&\triangleq&\lim_{N\rightarrow \infty}\lim_{n\rightarrow \infty}\frac{H(T^{n})}{nN}\nonumber\\
&=&R_{0}.
\end{eqnarray}

In addition, note that the encoding and decoding scheme for Theorem \ref{T5}
is exactly the same as that in \cite{St}, except that the transmitted message for receiver 1 is $s\oplus k^{*}$. Since receiver 1 knows $k^{*}$,
the decoding scheme for Theorem \ref{T5} is in fact the same as that in \cite{St}. Hence, we omit the proof of
$P_{e1}\leq \epsilon$ and $P_{e2}\leq \epsilon$ here.

It remains to show that $\lim_{N\rightarrow \infty}\Delta\geq R_{e}$, see the following.

Since the confidential message $S$ is encrypted by $S\oplus K^{*}$, the equivocation about $S$ is equivalent to the equivocation about $K^{*}$.
There are two ways for receiver 2 to obtain the secret key $k^{*}$. One way is that he tries to guess the $k^{*}$ from its alphabet $\mathcal{S}$.
The other way is that he tries to guess the feedback $y^{N}$ ($y^{N}$ is the output of channel 1 for the previous block,
and $k^{*}=g_{f}(y^{N})$) from
the conditional typical set $T^{N}_{[Y|Z]}(\delta)$, and this is because for a given $z^{N}$ and sufficiently large $N$,
$Pr\{(y^{N}\notin T^{N}_{[Y|Z]}(\delta)\}\rightarrow 0$. Note that there are $2^{NH(Y|Z)}$ sequences $y^{N}\in T^{N}_{[Y|Z]}(\delta)$
when $N\rightarrow \infty$ and $\delta\rightarrow 0$. Therefore, the equivocation about $S$ is
$\min\{\frac{\log \|\mathcal{S}\|}{N}=R_{1}, H(Y|Z)\}$, and note that $R_{1}\geq R_{e}$ and $H(Y|Z)\geq R_{e}$,
then $\lim_{N\rightarrow \infty}\Delta\geq R_{e}$ is obtained.

The details about the proof are as follows.

First, we will show that $K_{i}^{*}\oplus S_{i}$ is independent of $K_{i}^{*}$ and $S_{i}$, and this is used
in the proof of $\lim_{N\rightarrow \infty}\Delta\geq R_{e}$.

Since
$K_{i}^{*}$ is independent of $S_{i}$ ($2\leq i\leq n$), and all of them are uniformly distributed over $\mathcal{S}$,
the fact that $K^{*}\oplus S_{i}$ is independent of $K^{*}$ and $S_{i}$ is proved by the following (\ref{e4xx.11x}) and (\ref{e4xx.12x}).
\begin{eqnarray}\label{e4xx.11x}
Pr\{K_{i}^{*}\oplus S_{i}=a\}&=&\sum_{k_{i}^{*}\in \mathcal{S}}Pr\{K_{i}^{*}\oplus S_{i}=a, K_{i}^{*}=k_{i}^{*}\}\nonumber\\
&=&\sum_{k_{i}^{*}\in \mathcal{S}}Pr\{S_{i}=a\ominus k_{i}^{*}, K_{i}^{*}=k_{i}^{*}\}\nonumber\\
&=&\sum_{k_{i}^{*}\in \mathcal{S}}Pr\{S_{i}=a\ominus k_{i}^{*}\}Pr\{K_{i}^{*}=k_{i}^{*}\}\nonumber\\
&=&\frac{1}{\|\mathcal{S}\|}.
\end{eqnarray}
\begin{eqnarray}\label{e4xx.12x}
Pr\{K_{i}^{*}\oplus S_{i}=a, K_{i}^{*}=k_{i}^{*}\}&=&Pr\{S_{i}=a\ominus k_{i}^{*}, K_{i}^{*}=k_{i}^{*}\}\nonumber\\
&=&Pr\{S_{i}=a\ominus k_{i}^{*}\}Pr\{K_{i}^{*}=k_{i}^{*}\}\nonumber\\
&=&\frac{1}{\|\mathcal{S}\|^{2}}.
\end{eqnarray}

Then, $\lim_{N\rightarrow \infty}\Delta\geq R_{e}$ is proved by the following (\ref{e49abcnimei}).

\begin{eqnarray}\label{e49abcnimei}
\lim_{N\rightarrow \infty}\Delta&\triangleq&\lim_{N\rightarrow \infty}\lim_{n\rightarrow \infty}\frac{H(S^{n}|Z^{n})}{nN}\nonumber\\
&=&\lim_{N\rightarrow \infty}\lim_{n\rightarrow \infty}\frac{\sum_{i=2}^{n}H(S_{i}|S^{i-1},Z^{n})}{nN}\nonumber\\
&=^{(a)}&\lim_{N\rightarrow \infty}\lim_{n\rightarrow \infty}\frac{\sum_{i=2}^{n}H(S_{i}|\widetilde{Z}_{i},\widetilde{Z}_{i-1})}{nN}\nonumber\\
&\geq&\lim_{N\rightarrow \infty}\lim_{n\rightarrow \infty}\frac{\sum_{i=2}^{n}H(S_{i}|\widetilde{Z}_{i},\widetilde{Z}_{i-1},S_{i}\oplus K^{*}_{i})}{nN}\nonumber\\
&=^{(b)}&\lim_{N\rightarrow \infty}\lim_{n\rightarrow \infty}\frac{\sum_{i=2}^{n}H(S_{i}|\widetilde{Z}_{i-1},S_{i}\oplus K^{*}_{i})}{nN}\nonumber\\
&=&\lim_{N\rightarrow \infty}\lim_{n\rightarrow \infty}\frac{\sum_{i=2}^{n}H(K^{*}_{i}|\widetilde{Z}_{i-1},S_{i}\oplus K^{*}_{i})}{nN}\nonumber\\
&=^{(c)}&\lim_{N\rightarrow \infty}\lim_{n\rightarrow \infty}\frac{\sum_{i=2}^{n}H(K^{*}_{i}|\widetilde{Z}_{i-1})}{nN}\nonumber\\
&=^{(d)}&\lim_{N\rightarrow \infty}\lim_{n\rightarrow \infty}\frac{\sum_{i=2}^{n}\min\{NH(Y|Z),NR_{1}\}}{nN}\nonumber\\
&=&\lim_{N\rightarrow \infty}\lim_{n\rightarrow \infty}\frac{(n-1)\min\{NR_{1},NH(Y|Z)\}}{nN}\nonumber\\
&=&\min\{R_{1},H(Y|Z)\}\nonumber\\
&\geq&R_{e},
\end{eqnarray}
where (a) is from $S_{i}\rightarrow (\widetilde{Z}_{i},\widetilde{Z}_{i-1})\rightarrow (S^{i-1},\widetilde{Z}^{i-2},\widetilde{Z}^{n}_{i+1})$ is a Markov chain,
(b) is from
$S_{i}\rightarrow (S_{i}\oplus K^{*}_{i}, \widetilde{Z}_{i-1})\rightarrow \widetilde{Z}_{i}$ is a Markov chain, (c) follows from the fact that
$K^{*}_{i}\oplus S_{i}$ is independent of $K^{*}_{i}$ and $\widetilde{Z}_{i-1}$, and (d) is from
the fact that receiver 2 can guess the specific vector $\widetilde{Y}_{i-1}$ (corresponding to the key $K^{*}_{i}$) from the conditional typical set
$T^{N}_{[Y|Z]}(\delta)$, and  $K^{*}_{i}$
is uniformly distributed over $\mathcal{S}$ ($K^{*}_{i}$ is the key used in block $i$).

Thus, $\lim_{N\rightarrow \infty}\Delta\geq R_{e}$ is proved, and the proof of Theorem \ref{T5} is completed.

\section{Proof of Theorem \ref{T6}\label{appen7}}

In this section, we prove Theorem \ref{T6}: all the achievable $(R_{0}, R_{1}, R_{e})$ triples are
contained in the set $\mathcal{R}^{(nfo)}$.
Suppose $(R_{0}, R_{1}, R_{e})$ is achievable, i.e., for any given $\epsilon>
0$, there exists a channel encoder-decoder $(N, \Delta, P_{e1}, P_{e2})$ such that
$$\lim_{N\rightarrow \infty}\frac{\log\parallel \mathcal{T}\parallel}{N}= R_{0}, \lim_{N\rightarrow \infty}\frac{\log\parallel \mathcal{S}\parallel}{N}= R_{1},
\lim_{N\rightarrow \infty}\Delta\geq R_{e}, P_{e1}\leq \epsilon, P_{e2}\leq \epsilon.$$
Then we will show the existence of random variables $(U, K, A)\rightarrow (X, V)\rightarrow Y\rightarrow Z$ such that
\begin{equation}\label{e301pp}
0\leq R_{e}\leq R_{1},
\end{equation}
\begin{equation}\label{e302pp}
R_{0}\leq I(U;Z)-I(U;V),
\end{equation}
\begin{equation}\label{e303.1pp}
R_{1}\leq I(K;Y|U,A)-I(K;V|U,A),
\end{equation}
\begin{equation}\label{e303.2pp}
R_{0}+R_{1}\leq I(U,K,A;Y)-I(U,K,A;V),
\end{equation}
\begin{equation}\label{e304xxpp}
R_{e}\leq H(Y|Z).
\end{equation}

The proof of (\ref{e301pp}), (\ref{e302pp}), (\ref{e303.1pp}) and (\ref{e303.2pp}) are the same as those in the proof of Theorem \ref{T2},
and therefore, we omit it here. It remains to prove (\ref{e304xxpp}), see the following.

\begin{eqnarray}\label{e314xxpp}
\frac{1}{N}H(S|Z^{N})&\leq^{(a)}&\frac{1}{N}H(S|Z^{N})+\frac{1}{N}\delta(P_{e1})-\frac{1}{N}H(S|Y^{N},Z^{N})\nonumber\\
&=&\frac{1}{N}I(S;Y^{N}|Z^{N})+\frac{1}{N}\delta(P_{e1})\nonumber\\
&\leq&\frac{1}{N}H(Y^{N}|Z^{N})+\frac{1}{N}\delta(P_{e1})\nonumber\\
&\leq&\frac{1}{N}\sum_{i=1}^{N}H(Y_{i}|Z_{i})+\frac{1}{N}\delta(P_{e1})\nonumber\\
&=^{(b)}&\frac{1}{N}\sum_{i=1}^{N}H(Y_{i}|Z_{i},J=i)+\frac{1}{N}\delta(P_{e1})\nonumber\\
&=&H(Y_{J}|Z_{J},J)+\frac{1}{N}\delta(P_{e1})\nonumber\\
&=^{(c)}&H(Y|Z,J)+\frac{1}{N}\delta(P_{e1})\nonumber\\
&\leq&H(Y_|Z)+\frac{1}{N}\delta(\epsilon),
\end{eqnarray}
where (a) is from Fano's inequality, (b) is from $J$ is independent of $Y^{N}$ and $Z^{N}$, and (c) is from
$Y\triangleq Y_{J}$ and $Z\triangleq Z_{J}$.

By using $\lim_{N\rightarrow \infty}\Delta\geq R_{e}$, $\Delta=\frac{1}{N}H(S|Z^{N})$, (\ref{e314xxpp})
and letting $\epsilon\rightarrow 0$, the formula (\ref{e304xxpp}) is proved.

The proof of Theorem \ref{T6} is completed.

\section{Proof of Theorem \ref{T7}\label{appen8}}

\subsection{Converse Part of Theorem \ref{T7}\label{appen8.1}}

The converse proof of Theorem \ref{T7} is exactly the same as the proof of Theorem \ref{T4}, except that
$R_{e}\leq H(Y|Z)$. Note that the proof of $R_{e}\leq H(Y|Z)$ is the same as that in the proof of Theorem \ref{T6},
and hence we omit the proof here.

\subsection{Direct Part of Theorem \ref{T7}\label{appen8.2}}

In this section, we will show that any triple
$(R_{0},R_{1},R_{e})\in \mathcal{R}^{cf}$ is achievable.

\subsubsection{Code Construction\label{app8.2.1}}

Given a triple $(R_{0},R_{1},R_{e})$, choose a joint probability mass function $p_{U,K,V,X,Y,Z}(u,k,v,x,y,z)$
such that
$$0\leq R_{e}\leq R_{1},$$
$$R_{0}\leq I(U;Z),$$
$$R_{1}\leq I(K;Y|U),$$
$$R_{e}\leq H(Y|Z).$$

The confidential message set $\mathcal{S}$ and the common message set $\mathcal{T}$ satisfy the following conditions:
\begin{equation}\label{eapp1aa}
\lim_{N\rightarrow \infty}\frac{1}{N}\log\parallel \mathcal{S}\parallel=R_{1}=I(K;Y|U)-\gamma_{1},
\end{equation}
\begin{equation}\label{eapp2aa}
\lim_{N\rightarrow \infty}\frac{1}{N}\log\parallel \mathcal{T}\parallel= R_{0}=I(U;Z)-\gamma,
\end{equation}
where $\gamma$ and $\gamma_{1}$ are fixed positive numbers.
Let $\mathcal{S}=\{1,2,...,2^{NR_{1}}\}$ and $\mathcal{T}=\{1,2,...,2^{NR_{0}}\}$.

We use the block Markov coding method. The random vectors $U^{N}$, $K^{N}$, $V^{N}$, $X^{N}$, $Y^{N}$
and $Z^{N}$ consist of $n$ blocks of length $N$.
The common message for $n$ blocks is $T^{n}\triangleq (T_{1},...,T_{n})$,
which is composed of $n$ i.i.d. random variables uniformly distributed over $\mathcal{T}$.
The confidential message for $n$ blocks is $S^{n}\triangleq (S_{2}, S_{3},...,S_{n})$,
where $S_{i}$ ($2\leq i\leq n$) are i.i.d. random variables uniformly distributed over $\mathcal{S}$.
Note that in the first block, there is no $S_{1}$.

Let $\widetilde{Z}_{i}$ ($1\leq i\leq n$) be the output of channel 2 for block $i$, $Z^{n}=(\widetilde{Z}_{1},...,\widetilde{Z}_{n})$,
$Z^{\overline{j}}=(\widetilde{Z}_{1},...,\widetilde{Z}_{j-1},\widetilde{Z}_{j+1},...,\widetilde{Z}_{n})$ $(1\leq j\leq n)$.
Similarly,  $Y^{n}=(\widetilde{Y}_{1},...,\widetilde{Y}_{n})$, and
$\widetilde{Y}_{i}$ ($1\leq i\leq n$) is the output of channel 1 for block $i$.
The specific values of the above random vectors are denoted by lower case letters.

Code-book generation:

\begin{itemize}

\item \textbf{(Construction of $U^{N}$)}

The construction of $U^{N}$ for each block is exactly the same as that in the proof of Theorem \ref{T3}, see the following.

For the $i$-th block ($1\leq i\leq n$), given a common message $t_{i}$ ($t_{i}\in \mathcal{T}$), generate a corresponding
$u^{N}(t_{i})$ i.i.d. according to the probability mass function $p_{U}(u)$.

\item \textbf{(Construction of $K^{N}$)}

In the first block, for a given $u^{N}(t_{1})$, generate a
corresponding sequence $k^{N}$ i.i.d.
according to the probability mass function $p_{K|U}(k_{i}|u_{i})$.

In the $i$-th block ($2\leq i\leq n$), firstly we generate a mapping
$g_{f}: \mathcal{Y}^{N}\rightarrow \mathcal{S}$. Define a random variable $K_{i}^{*}=g_{f}(\widetilde{Y}_{i-1})$ ($2\leq i\leq n$),
which is uniformly
distributed over $\mathcal{S}$, and $K_{i}^{*}$ is independent of $S_{i}$.
Reveal the mapping $g_{f}$ to both receivers and the transmitter.
Then, when the transmitter receives the output $\widetilde{y}_{i-1}$ of the $i$-1-th block, he computes
$k_{i}^{*}=g_{f}(\widetilde{y}_{i-1})\in \mathcal{S}$.

Given the transmitted sequence $u^{N}(t_{i})$ and the encrypted confidential message $s_{i}\oplus k_{i}^{*}$
(where $\oplus$ is the modulo addition over $\mathcal{S}$), generate a
corresponding sequence $k^{N}$ i.i.d.
according to the probability mass function $p_{K|U}(k_{i}|u_{i})$. Index $k^{N}$ by $s_{i}\oplus k_{i}^{*}\in \mathcal{S}$.

\item \textbf{(Construction of $X^{N}$)}

For each block, the $x^{N}$ is generated
according to a new discrete memoryless channel (DMC) with inputs $k^{N}$, $u^{N}$, $v^{N}$,
and output $x^{N}$. The transition probability of this new DMC is $p_{X|U,K,V}(x|u,k,v)$.
The probability $p_{X^{N}|U^{N},K^{N},V^{N}}(x^{N}|u^{N},k^{N},v^{N})$ is calculated as follows.
\begin{equation}\label{eapp4.5xxxx}
p_{X^{N}|U^{N},K^{N},V^{N}}(x^{N}|u^{N},k^{N},v^{N})=
\prod_{i=1}^{N}p_{X|U,K,V}(x_{i}|u_{i},k_{i},v_{i}).
\end{equation}

\end {itemize}

Decoding:

Receiver 2: For each block, given a vector $z^{N}\in \mathcal{Z}^{N}$, try to find a sequence $u^{N}(\hat{t})$ such that
$(u^{N}(\hat{t}),z^{N})\in T^{N}_{UZ}(\epsilon_{3})$.
If there exists a unique sequence, put out the corresponding $\hat{t}$.
Otherwise, declare a decoding error.

Receiver 1: For the first block,  given a vector $\widetilde{y}_{1}\in \mathcal{Y}^{N}$, try to find a sequence $u^{N}$ such that
$(u^{N},\widetilde{y}_{1})\in T^{N}_{UY}(\epsilon_{4})$.
If such a sequence does not exist, or there are more than one such sequence, declare a decoding error.
Denote the corresponding sequence by $u^{N}(\hat{t}_{1})$, put out the corresponding index $\hat{t}_{1}$.

For block $i$ ($2\leq i\leq n$), given a vector $\widetilde{y}_{i}\in \mathcal{Y}^{N}$, try to find a sequence $u^{N}$ such that
$(u^{N},\widetilde{y}_{i})\in T^{N}_{UY}(\epsilon_{4})$.
If such a sequence does not exist, or there are more than one such sequence, declare a decoding error.
Denote the corresponding sequence by $u^{N}(\hat{t}_{i})$, put out the corresponding index $\hat{t}_{i}$.

After decoding $u^{N}(\hat{t}_{i})$ and $\hat{t}_{i}$, try to find a sequence $k^{N}(u^{N}(\hat{t}_{i}))$ such that
$(k^{N}(u^{N}(\hat{t}_{i})),\widetilde{y}_{i})\in T^{N}_{KY|U}(\epsilon_{5})$.
If there exists a sequence, put out the corresponding index $\hat{s}_{i}\oplus k_{i}^{*}$.
Otherwise, declare a decoding error. Finally, since receiver 1 knows $k_{i}^{*}$ ($k_{i}^{*}=g_{f}(\widetilde{y}_{i-1})$),
put out the corresponding $\hat{s}_{i}$ from $\hat{s}_{i}\oplus k_{i}^{*}$.

\subsubsection{Proof of Achievability\label{app8.2.2}}

Note that the encoding and decoding scheme for Theorem \ref{T7}
is exactly the same as that in \cite{St}, except that the transmitted message for receiver 1 is $s\oplus k^{*}$. Since receiver 1 knows $k^{*}$,
the decoding scheme for Theorem \ref{T7} is in fact the same as that in \cite{St}. Hence, we omit the proof of
$P_{e1}\leq \epsilon$ and $P_{e2}\leq \epsilon$ here.

It remains to show that $\lim_{N\rightarrow \infty}\Delta\geq R_{e}$, see the following.

Since the confidential message $S$ is encrypted by $S\oplus K^{*}$, the equivocation about $S$ is equivalent to the equivocation about $K^{*}$.
There are two ways for receiver 2 to obtain the secret key $k^{*}$. One way is that he tries to guess the $k^{*}$ from its alphabet $\mathcal{S}$.
The other way is that he tries to guess the feedback $y^{N}$ ($y^{N}$ is the output of channel 1 for the previous block,
and $k^{*}=g_{f}(y^{N})$) from
the conditional typical set $T^{N}_{[Y|Z]}(\delta)$, and this is because for a given $z^{N}$ and sufficiently large $N$,
$Pr\{(y^{N}\notin T^{N}_{[Y|Z]}(\delta)\}\rightarrow 0$. Note that there are $2^{NH(Y|Z)}$ sequences $y^{N}\in T^{N}_{[Y|Z]}(\delta)$
when $N\rightarrow \infty$ and $\delta\rightarrow 0$. Therefore, the equivocation about $S$ is
$\min\{\frac{\log \|\mathcal{S}\|}{N}=R_{1}, H(Y|Z)\}$, and note that $R_{1}\geq R_{e}$ and $H(Y|Z)\geq R_{e}$,
then $\lim_{N\rightarrow \infty}\Delta\geq R_{e}$ is obtained.

The detail about the proof of $\lim_{N\rightarrow \infty}\Delta\geq R_{e}$ is exactly the same as (\ref{e49abcnimei}),
and it is omitted here.

Thus, $\lim_{N\rightarrow \infty}\Delta\geq R_{e}$ is proved, and the direct part of Theorem \ref{T7} is completed.

The proof of Theorem \ref{T7} is completed.

\end{document}